\documentstyle[11pt,aaspp4]{article}
\def\etal{{\it et al.\ }}

\begin{document}
\title{{\huge Normal Galaxies} \\ \vspace{5mm}  INES Guide No. 2 
\\ International Ultraviolet Explorer IUE Newly 
Extracted Spectra}

\author{Liliana Formiggini{\footnote{Email: lili@wise.tau.ac.il}}}

\author{Noah Brosch{\footnote{Email: noah@wise.tau.ac.il}}} 

\affil{The Wise Observatory and 
the School of Physics and Astronomy,
Raymond and Beverly Sackler Faculty of Exact Sciences,
Tel Aviv University, Tel Aviv 69978, Israel}


\begin{abstract}
This guide presents, in a uniform manner, all the information collected by the
IUE satellite on normal galaxies.  It contains information on
274 galaxies and it supersedes the previous IUE guide to normal galaxies (Longo
\& Capaccioli 1992). The data shown here are restricted to
galaxies defined as ``normal'' by the observer, and entered as such in the IUE
data file header. The information is also restricted to low-dispersion spectra
obtained through the large apertures of IUE.
For the first time, we provide spectral information from {\bf well-defined}
and identifiable
locations in the target galaxies. These are mostly located close to the photocenter
of each object, although there are few exceptions. Each representative spectrum 
of a galaxy consists of  a short-wave (SW) and a long-wave 
(LW) IUE low-dispersion
spectrum (where available) combined into a single spectrum covering the 
wavelength range 1150\AA\, to 3350\AA\,. We selected the two spectra to
be combined so as to be, preferably,
the deepest exposures available in the INES archive. Each representative spectrum 
is accompanied by two images of the galaxy, on which the locations
of the SW entrance aperture and the LW entrance aperture are marked.

\end{abstract}

\keywords{UV, galaxies, spectroscopy  }
\newpage

\section{Foreword: The INES Access Guides}

The International Ultraviolet Explorer (IUE) Satellite project was a joint effort 
between NASA, ESA and the PPARC. The IUE spacecraft and instruments were 
operated in a Guest Observer mode to allow ultraviolet spectrophotometry at two 
resolutions in the wavelength range from  1150\AA\, to 3200\AA\ : low resolution 
$\frac{\lambda}{\Delta\lambda}$=300 
($\sim$1,000 km/sec.) and a high resolution mode 
$\frac{\lambda}{\Delta\lambda}$=10,000 ($\sim$19 km/sec.). The IUE 
spacecraft, its scientific instruments as well as the data acquisition and reduction 
procedures, have been described in ``Exploring the Universe with the IUE Satellite", 
Part I, Part VI and Part VII (Astrophysics and Space Sciences Library volume 129, Y. 
Kondo, Editor-in-Chief, Kluwer Acad. Publ. Co.) and references therein. A more recent 
overview of the IUE project is given in the conference proceedings of the last IUE 
conference ``Ultraviolet Astrophysics beyond the IUE Final Archive''  (ESA SP-413, 
1998, Eds. W.Wamsteker and R. Gonzalez Riestra) and in ``IUE Spacecraft Operations 
Final Report'' (ESA SP-1215, 1997, A. Perez Calpena \& J.Pepoy).  Additional  
information on the IUE project can also be found at URL: 
http://www.vilspa.esa.es/iue/iue.html. 

From the very beginning of the project (launched on 26 January 1978), it was 
expected that the archival value of the data obtained with IUE would be very high. This 
expectation has been borne out fully after 18.6 years of orbital operations (on 30 
September 1996 the science operations with the IUE spacecraft were stopped). 
The average IUE Archive data retrieval rate, during the operational phase of the project, 
has been some 61,000 spectra per year. This compares to a new data access rate of 
5,500 spectra per year. Considering that the demand for observing time continued to 
exceed the available time by a factor of three until the end of science 
operations, it is clear 
that the IUE Archive remains an important source of data. The IUE ULDA/USSP 
(Uniform Low Dispersion Archive/ULDA Support Software Package) was developed 
by ESA in the mid-eighties (Wamsteker et al., 1989, Astron. \& Astrophys. Suppl. Ser., 
{\bf 79}, pg. 1-10) as the first astronomical archive with direct access for users on a 
world-wide basis. Over the ten years that the ULDA has been supporting  
IUE data retrieval, 
it has driven more than 50\% of all IUE Archive usage. The quantity of data in the IUE 
Archive is sufficiently large that it is not necessarily simple to address the data 
efficiently in the context of an astrophysical problem, even  though access to the data is 
extremely easy. Therefore, the series of INES Access Guides is intended to facilitate the 
use of the IUE Archive for scientists with a specific astrophysical problem in mind.

The extremely good reception of the distributed archive model by the scientific 
community has led to the decision to develop the Final Archive server with a similar 
philosophy, in the form of INES (IUE Newly Extracted Spectra). The INES system is  
a complete system design, with 
\begin{itemize}
\item improvements to the data 
\item a structure to facilitate the direct application to scientific analysis, and 
\item an integrated and distributed data retrieval system. 
\end{itemize}

Detailed information on the INES System and its data content has been published 
in a series of papers ``The INES System''. The specific references are:
Rodriguez-Pascual \etal 1999, Astron. \& Astrophys. Suppl. Ser., {\bf 139}, pg.183-198; 
Cassatella \etal 2000, Astron. \& Astrophys. Suppl. Ser., {\bf 141}, pg.331-342; 
Gonzalez-Riestra \etal 2000 Astron. \& Astrophys. Suppl. Ser, {\bf 141}, pg.343-356; 
Wamsteker \etal 2000, Astroph. \& Space Sci., in press. 
The INES Users Guide, also collecting these publications, has been published by ESA 
Publications Division in the INES Newsletter (March 2000).

 At  the time of this writing 19 National Hosts have functional installations  of the INES 
system  Version 2.0 (see below), giving to the end user, direct access to all 104,000 IUE spectra, 
with the Principal Center (LAEFF) at URL  http://ines.vilspa.esa.es/ines/  
and its Mirror Site for North America at the Canadian Astronomy Data Centre (CADC) 
http://ines.hia.nrc.ca in Victoria. The ESA involvement in the IUE Project activities 
will come to an end in 2000, and, from that moment onward, the INES system will 
become a part of the astronomical heritage of the IUE project. More National Hosts are 
forseen to come on-line in the future.
The series of INES Access Guides is forseen to continue and consists of a number of 
subject-oriented books, for which a specialist in the field has been invited to take the 
scientific responsibility. INES  Access Guide No. 2 treats the data for Normal Galaxies 
and has been compiled by  Liliana Formiggini and Noah Brosch of the Wise 
Observatory and the School of Physics and Astronomy, Tel Aviv University. This 
volume supersedes the ULDA Guide No. 3.
Further volumes of INES Access Guides  will be published whenever the necessary 
data compilation has been completed by the authors. The list of previously published 
ULDA \& INES Access Guides is given below, as well as the INES Guides currently in 
preparation.
For details of the access to INES through the National Hosts we refer to the 
information supplied at http://ines.vilspa.esa.es/ines/, or recommend contacting the 
INES Helpdesk at LAEFF at VILSPA, Madrid, Spain ineshelp@iuearc.vilspa.esa.es. 
Other queries about the data or any specific questions about data content in relation to 
the INES system should also be directed there or to the National Host Institutes.

\hspace{10cm}    Willem Wamsteker

\subsection{Previously issued IUE-ULDA Access Guides:}
\begin{description}
\item No. 1  ESA SP-1114    C. la Dous  
{\bf Dwarf Novae and Nova-like Stars}.
\item No. 2  ESA SP-1134    M. Festou  
{\bf Comets}.
\item No. 3  ESA SP-1146    G. Longo, M. Capaccioli  
{\bf Normal Galaxies}.  
\item No. 4  ESA SP-1153 (Vol. A \& B)  T.J.-L. Courvoisier, S. Paltani  
{\bf Active Galactic Nuclei}. 
\item No. 5  ESA SP-1181 (Vol. I \& II) C. la Dous, A. Gimenez
{\bf Chromospherically Active Binary Stars}.
\item No. 6  ESA SP-1189    E. Cappellaro, M. Turatto, J. Fernley
{\bf Supernovae}
\item No. 7  ESA SP-1203     M. Franchini, C. Morossi, M.L. Malagnini
{\bf K Stars}
\item No. 8  ESA SP-1205    A.I. Gomez de Castro, M. Franqueira
{\bf T Tauri Stars}
\end{description}

\subsection{INES Guides:}
\begin{description}
\item No. 1  ESA SP-1237    A.I. Gomez de Castro, A. Robles
{\bf Herbig-Haro Objects}   
\item No. 2  ESA SP-1239 L. Formiggini,  N. Brosch
{\bf Normal Galaxies} supersedes ULDA Guide No.3
\end{description}

\subsection{INES Guides in preparation:}
\begin{description}
\item No. 3    A. Niedzielski  {\bf Wolf-Rayet Stars} 
\item No. 4    M. Festou    {\bf Comets}   
    Supersedes ULDA Guide \# 2
\item No. 5   A. Cassatella \& R. Gonzalez-Riestra  {\bf Novae}
\end{description}

\subsection{INES National Hosts (as of 18 February 2000):}
\begin{tabular}{c|c}
         Country         &               Institution \\ \hline
            Argentina     &    Facultad de Ciencias  Astronomicas y Geofisicas, \\
                         &               Buenos Aires \\
              Austria      &      Kuffner Observatory, Vienna \\
              Belgium      &      Royal Observatory of  Belgium, Ukkel \\
              Brazil       &      Instituto Astronomico e  Geofisico, Sao Paulo \\
                Canada     &        Canadian Astronomy Data Center, Victoria \\
                          &           (Principal   Centre Mirror Site) \\
              China       &       Center for Astrophysics, \\
                           &             Hefei (local access only) \\
              ESO          &    European Southern  Observatory, Garching \\
                            &            (local access only) \\
              Israel       &      Wise Observatory, Tel Aviv University, Tel Aviv \\
              France      &       Centre de Donees Astronomiques, Strasbourg \\
              Italy       &       Astronomical Observatory,     Trieste \\
              Japan        &      National Astronomical  Observatory, Tokyo \\
              Korea       &       Chungbuk National University, Cheongju \\
              The Netherlands  &  Astronomical Institute,  Utrecht \\
              Poland       &      Center for Astronomy, Torun \\
               Russia      &    Institute of Astronomy of  the Russian Academy of \\
                          &              Sciences, Moscow \\
              Spain      &        LAEFF, Madrid (Principal Centre) \\
              Sweden      &      Astronomical Observatory Uppsala \\
            &       (serving the Nordic  Countries) \\
              Taiwan       &      Institute of Physics and Astronomy, Chung-Li \\
              United Kingdom  &   Rutherford Appleton Laboratory, Didcot \\
              United States   &   Space Telescope Science   Institute, Baltimore
\end{tabular}

\newpage

\section{Introduction}
 
    One of the greatest successes ever of any orbiting astronomical 
    instrument lies with the IUE (International Ultraviolet Explorer)
observatory. Launched on 26 January 1978 
    for a nominal three-year mission, the observatory operated for 18.7 
    years yielding more than 110,000 spectra of nearly 10,000 diverse 
    astronomical targets. IUE operations closed down on 30 September 1996. 
The International Ultraviolet Explorer satellite was a joint project
between NASA, ESA and PPARC (formerly SERC in UK). This was  a trilateral project, 
in which  NASA  provided the spacecraft,
telescope, spectrographs and one ground station, ESA the solar panels
and the second ground station, and the UK supplied the four spectrograph detectors.
In addition to controlling the satellite, the ground stations acted as typical
astronomical observatories, except that the telescope was in a high
orbit around the Earth. ESA's IUE Observatory was established in 1977 at the Villafranca
del Castillo
Satellite Tracking Station (VILSPA), Villanueva de la Ca\~{n}ada, Madrid, Spain.
The NASA IUE Observatory was located in the Goddard Space Flight Centre (GSFC),
in Greenbelt, MD.

    The experiment consisted of a 45-cm diameter f/15 Ritchey-Chr\'{e}tien 
Cassegrain telescope with an image quality of two arcsec feeding 
    two alternative spectrometers, one 
    for the range 1150-1980\AA\, (short-wave=SW) and the other for 1800-3350\AA\,
(long-wave=LW). Two different 
    dispersions were available, low (R=270 at 1500\AA\, and 400 at 2700\AA\,) 
  and high (R=1.8 10$^4$ at 1400\AA\, and 1.3 10$^4$ at 2600\AA\,, using an 
    echelle arrangement). The echelle observations used 3 arcsec circular
apertures; the low dispersion observations used mainly the oval apertures
with dimensions 10"$\times$20".
Each of the two spectrographs was equipped with two
cameras, a prime one (P) and a redundant one (R). The sensitivity of IUE, 
for low dispersion operations, was as follows:

\vspace{5mm}

\begin{tabular}{cc}
Camera (spectral range) & Sensitivity  \\ \hline
SWP (1150\AA\,-1980\AA\,) &  2 10$^{-15}$ erg s$^{-1}$ cm$^{-2}$ \AA\,$^{-1}$\\
LWP (1850\AA\,-3350\AA\,) &  1 10$^{-15}$ erg s$^{-1}$ cm$^{-2}$ \AA\,$^{-1}$\\
LWR (1850\AA\,-3350\AA\,) &  2 10$^{-15}$ erg s$^{-1}$ cm$^{-2}$ \AA\,$^{-1}$\\
SWR & Never operational \\
\end{tabular}

\vspace{5mm}

In addition, two Fine Error Sensors (FES) were incorporated. These transmitted
an optical video image of the sky area around the target reflected off of 
the focal plane, to the controlling ground station. 
FES-1 was never used in operations. FES-2 was always used for fine guidance
throughout the mission. It has also been important as a photometer to
measure the optical brightness of the observed sources. In 1991, scattered
light entering the telescope required a revision of the guidance procedures and
affected the photometric performance of the FES.

The only serious problems encountered with IUE during its operation stemmed 
from the failures of five of the six
gyroscopes in its attitude control system (in the years 1979, 1982, 1982, 1985, and 1996).
When the fourth gyroscope failed, IUE continued operations thanks to an
innovative reworking of its attitude control system by using the fine sun
sensor as a gyro substitute. Even with another gyroscope lost in the last year of
operation, IUE could
still be stabilized in three-axes with a single gyroscope, by adding
star-tracker measurements.
    During the unexpectedly long operation period of IUE, its NASA operators 
    at GSFC and ESA personnel at VILSPA devised work-around methods to  operate, despite 
    unexpectedly high levels of stray light, when a piece of thermal blanket 
    or reflecting tape fluttered in front of the telescope aperture in 1991. 

Until October 1995, IUE was in continuous operation, controlled for 16 hours
daily from GSFC and for the remaining eight hours from VILSPA. After that, ESA took
on a major role to alleviate financial problems of the partners. The
operational schemes were completely redesigned and an innovative control
system was implemented. With these innovations, it became feasible to perform
science operations controlled only from VILSPA. For practical reasons, only
16 hours were used for scientific operations, while the eight hours in the 
low-quality (high radiation background) part of the orbit were used for 
spacecraft housekeeping. In
February 1996, ESA decided to discontinue the satellite operations. IUE remained
operational until 30 September 1996, when its remaining hydrazine fuel was deliberately
vented, its batteries were drained, and its transmitter was turned off.

\subsection{The archive}

The stability of the IUE instrument, and the care taken to ensure a proper
calibration throughout the mission, make the IUE database a very valuable
archival resource for specific investigations in the domain of astronomical
ultraviolet spectroscopy.

The IUE Uniform Low-Dispersion Archive (ULDA) was the first attempt to put the
results of IUE at the disposition of the astronomical community in an orderly manner.
Version 3.0 of the ULDA was released in September 1990 and contained 98.7\% of all
the low-resolution spectra obtained by IUE before January 1, 1989 ($\sim$44,000 spectra).
The ULDA was installed at a number of national hosts and was made available to
general users from these hosts.

 
    The lasting value of the IUE archive became even more evident 
    after the final reprocessing of all the low-dispersion spectra into the 
    final IUE Newly Extracted Spectral Archive (INES) was completed. 
The reprocessing was done using improved extraction techniques and
calibrations (NEWSIPS: Nichols \etal 1993).
The IUE Final Archive, another collaborative effort by NASA, ESA, and PPARC, was identified 
during the final phase of the mission as a necessary step in the production of a 
high-quality and uniform data base. This archive was seen as the final repository of the
information collected by IUE, produced at a stage when specialized knowledge of
the instrument and of its calibration procedures were still available. The final archive 
will be maintained accessible to the scientific community as the historical
reference of the IUE mission and as a source of information for future studies 
based on IUE UV spectra. It is a special tribute to the leadership of the IUE project
that most of the final archive production was done while the satellite was still
in operation.

The final version of the IUE archive 
    contains 104,471 spectra reprocessed with the most up-to-date calibration
    and includes high-resolution echelle spectra binned to the resolution of the low
    dispersion observations. The low-resolution version of the IUE archive 
has been installed at the ESA/INTA INES Primary Center 
at VILSPA/LAEFF and at 19 national or institutional hosts (as of 18 February 2000).

\subsection{Information content}

    Although the IUE data set does not represent a uniform survey of the 
    sky, the large variety of objects observed by it offers unique opportunities 
    to derive ``average'' properties of celestial populations. This has been used by 
    many ({\it e.g.} Fanelli {\it et al.} 1987) to derive UV-to-optical color 
    indices for various stellar spectral types and luminosity classes. These are 
    later used to derive transformations, to create models of the UV sky (Brosch 1991), 
    or to determine the level of the diffuse UV background. 
    Observations of galaxies were used to determine average UV spectra of 
    irregular, spiral, and elliptical galaxies, and of galactic bulges 
    (Ellis \etal 1982, Burstein {\it et al.} 1988, Kinney {\it et al.} 
    1993, Storchi-Bergmann 
    {\it et al.} 1994), important for the derivation of cosmological k-corrections, 
and for analyzing observed
properties of high-redshift clusters of galaxies (Steindling \etal 2000).

To help the logical usage of the information collected by IUE, atlases of UV 
spectra of selected types 
    of objects, based on IUE data were published ({\it i.e.} Longo \& 
    Capaccioli 1992 for normal galaxies, or Courvoisier \& Paltani 1992 for
active galaxies). Apart from these special-purpose atlases, note those 
    dedicated to the classification of stars from their UV spectra 
published by ESA and by NASA   
    (Heck \etal 1984; Wu \etal 1991). 

\subsection{Galaxies in the IUE data set}


The ULDA 
Access Guide No. 3 (Longo \& Capaccioli 1992) included spectra obtained
up to the 31st of December 1991. The availability of the IUE Final Archive, and of INES,
along with the additional information collected until the  cessation of IUE
observations on 30 September 1996, were the reasons to compile a new 
collection of UV spectra of normal galaxies, obtained by the IUE Observatory.
This collection is presented in this publication. We believe that its compilation will
further enhance the lasting value of the IUE observations.

\section{UV radiation from galaxies}

 Far UV radiation from galaxies was detected in the late sixties by space satellites
such as the OAO-2 and later the TD-1. It was the advent of the IUE however, that provided
a large number  of spectra for many galaxies, of different morphological
types, that confirmed and greatly expanded the earlier UV data. 
Attempts to explain the UV emission from galaxies led to a flourishing of modeling
work  and to new views on star formation and chemical evolution. 
In spite of the enormous contribution of the IUE experiment, the information about
UV emission from  galaxies is very sparse and the astronomical
   community  still lacks a large 
    sample of a few thousand galaxies with good UV information, from which to perform 
 adequate statistical studies. This information is very important in order to understand 
evolution in the Universe.

    Significant information on selected objects, mainly on stellar 
    populations and the nature of the ISM, was obtained from the early IUE spectra 
    (O'Connell 1992). Similar observational data, combining UV with optical 
    and near-IR spectrophotometry through matched apertures, were  
    used to derive template spectral energy distributions (SEDs) for various 
    types of galaxies (Storchi-Bergmann {\it et al.} 1994; McQuade {\it et 
    al.} 1995). A combination of IUE observations and data from other UV 
    imaging missions was used to extract ``total'' UV information on 
    galaxies (Longo \etal 1991; Rifatto \etal 1995a, 1995b). Some of the latter information
was included in the ULDA Guide to Normal Galaxies (Longo \& Capaccioli 1992) and
in a comparison of galaxy properties in the UV using synthetic photometry from the IUE
spectra in seven photometric bands, five of which matched the ANS bands (Longo \etal 1991).
 
    There is hope to derive the star-forming histories of galaxies through a 
    combination of data from the UV to the near-IR, in the manner of the 
    Storchi-Bergmann {\it et al.} (1994; SB2) templates. The significant UV data 
collected     by IUE from normal galaxies is usually at $\lambda>$1400\AA\,, except for 
the very young starburst galaxies and for some ellipticals. 
This spectral region, in late-type galaxies, contains mainly radiation from A-type 
    stars and requires extrapolation of the stellar population to earlier types
in order 
    to account for Lyman continuum photons. The flux below 1400\AA\,, observed in
starburst galaxies, originates from stars earlier than type A, mainly B stars, although 
sometimes a contribution by field O-stars cannot be ruled out (Brosch \etal 1999). Some 
elliptical galaxies exhibit upturns of their spectral energy distributions below
2000\AA\, (Burstein \etal 1988; Bonatto \etal 1996; O'Connell 1999). These are 
presumably produced by
low-mass, helium-burning stars on the extreme horizontal-branch and in later stages 
of evolution.

 It is impractical to rely only 
    on the detailed modeling of spectral features in the optical region
    in order to  understand large populations of galaxies in terms of stellar 
    populations and star-formation histories.
    One should combine 
    information from many spectral bands, covering as wide a spectral region  
    as possible.  
    In the absence of very deep UV surveys in more than a single spectral band, 
    such as those expected to result from the GALEX all-sky two-band UV survey 
(Bianchi \& Martin 1997), our information 
    about significant numbers of galaxies measured in the UV originates from the SCAP-2000 
    (Donas {\it et al.} 1987), FOCA (Milliard {\it et al.} 1992), and FAUST
(Deharveng \etal 1994; Bowyer \etal 1995; Brosch \etal 1995, 1997, 1999, {\it et seq.})
    measurements. These consist of integrated photometry at 1650\AA\, or 
2000\AA\, of a few 
    hundred galaxies. 
UV spectral information for about two dozen galaxies was also obtained by the ASTRON
observatory (Boyarchuk 1994).
In the 2000\AA\, band, and in the UV brightness range 
    16.5-18.5, galaxies apparently dominate the source counts (Milliard {\it et al.} 1992). The 
    corresponding blue magnitudes of these galaxies are B=18--20 mag. and their 
    typical color index is $[2000-V]\approx-$1.5. Comparing this color index with 
    the template spectra of Kinney {\it et al.} (1996), the FOCA-detected galaxies 
seem to
    fit the SB2 template, {\it i.e.,} a slightly reddened starburst galaxy. 
   The claim by 
    the FOCA group of a large contribution of UV galaxies in deep-UV sky
observations is supported by 
    theoretical arguments requiring a fast-evolving population of perhaps 
    dwarf galaxies for z=0.2-1.0, in order to explain the faint source counts in 
    other spectral domains (Ellis 1997). 
 
\section{The INES data}

The data sets provided by INES consist of low-resolution spectra extracted with an
improved method from the line-by-line images of the IUE Final Archive, and of high-resolution
spectra resampled to the low-resolution wavelength step. The new extraction of
the low-resolution spectra includes:

\begin{enumerate}

\item A better noise model (Schartel \& Rodriguez-Pascual 1998).

\item A better spectral extraction procedure, resulting in improved extraction of
spectra with lines and in a more adequate background estimation (Rodriguez-Pascual \etal 1998).

\item A better propagation of the quality flags, with a larger number of pixels
flagged and more correct information about potential problems.

\item A homogenized wavelength scale for the LW cameras.

\item A major improvement in removing the contribution of stray light into 
the LWP camera in images obtained after 1991.

\item A revision of the flux density scale.

\end{enumerate}

Another important aspect of INES {\it vs.} ULDA is the presentation of 
spectra in standard FITS format; this implies that the spectra can easily be
manipulated by the standard image-processing software packages in general use by the
astronomical community.

The noise model used for the production of INES is derived empirically, from hundreds of
flat-field images with different exposure times. This was required, given that the IUE detectors 
are television-type and are different in their behavior from  CCD detectors. An improved noise
model for the IUE spectra was a necessary first-step of an improved extraction procedure, 
because the determination of the spatial extent of the object spectrum in the cross-dispersion
direction is based on the signal-to-noise (S/N) ratio along the entrance aperture. Also, 
the determination of the errors in the
final extracted spectra is based on propagation of the errors from all the previous stages
of the extraction.

The background determination for INES is performed by
fits of Chebyshev polynomials across the IUE spectra external to the entrance aperture.
The polynomial fit is assumed to be constant across the aperture.
The extraction profile is determined by finding the extent of the spectrum through a
spline fit along the spatial direction, for extents of S/N$>$30 along the direction
of the dispersion, and with a minimum of seven wavelength steps. In cases of weak spectra, 
where the S/N is
below a certain threshold, the extraction is done by adding up all the flux within the
entire aperture. In many instances relevant to galaxies, when the aperture is much larger
than the dimensions of the target and the surface brightness is relatively low, 
this is the proper procedure.

The accuracy of INES flux extraction has been tested by Schartel \& Skillen (1998) and
was found that, in general, NEWSIPS and INES gave consistent results. The definite advantage of INES was demonstrated  for spectra with strong and narrow emission lines, where
INES gave consistently more reliable results.

In addition, the production of INES was done through a strict configuration control of
all the re-processing stages. This resulted in improved quality control of the spectra 
and also of the information
entered in the image headers. This included interactive verification of the
input parameters against the hand-written observing logs, as well as automatic checks of the validity
of values, correct sizes, and input formats of the raw images.

\section{Selection of galaxies for inclusion in the present guide}

The selection of objects to be included in this catalog is  based primarily on the
object classification included in the FITS header of the object spectrum in INES
and partly on object selection from the INES archive. The IUE observing class
is a two-digit code describing the observed target. In selecting the objects included here,
we chose codes 80 (spiral galaxy), 81 (elliptical galaxy), 82 (irregular
galaxy), and 88 (emission-line galaxy, non-Seyfert). Searches of the INES archive returned
370 objects corresponding to these definitions. Further consideration of 
additional data about these 370
objects, mainly collected from the NED\footnote{The NASA/IPAC Extragalactic 
Database (NED) is operated by the Jet Propulsion
Laboratory, California Institute of Technology, under contract with the National
Aeronautics and Space Administration.} data base, indicated that some of them harbor 
active galactic nuclei (AGNs). These objects may not have been known as 
AGNs at the time of the IUE observations,
or the Guest Observer (GO) did not identify the object as an AGN when the 
Resident Astronomer (RA)
entered the necessary information in the exposure header. Such objects were deleted from
the candidate list.

On the other hand, opposite cases of missed normal galaxies could exist. We did not search 
the more than 100,000 spectra  on INES for objects 
classified by the GO, and entered by the RA in the image header, as anything but 
classes 80, 81, 82, or 88,
which may be ``normal galaxies'' as defined here.
These objects are, therefore, excluded from this guide.

Another reason  for rejecting candidates selected as normal galaxies were a few 
cases where an object was identified as a normal galaxy, had a spectrum present in
the archive, but the extracted spectrum contained essentially no flux. In other
cases, the location of the IUE entrance aperture was patently off the object and
sampled only clear sky, yielding a zero net-flux spectrum in INES.
Such cases were not included in our compilation, but have a remark in the
appropriate section of the comments.

The 274 galaxies included in the present guide are listed in Table 1. We present there
the leading name of the galaxy as it appears in the homogeneous identification of IUE, 
its J2000 equatorial coordinates, its morphological type 
and its T-type, its total B magnitude, and (B--V) and (U--B) colors, as listed in the
LEDA database.

\section{Presentation of the guide data}

Given the large number of normal galaxies in the sample, we decided to present for 
each object the most relevant 
information  on a single page of this Guide. An individual page contains,
therefore, the
following information:

\begin{enumerate}

\item A header with the name of the galaxy as given in INES, its J2000 
celestial coordinates 
($\alpha$ and $\delta$), and its Galactic coordinates (l and b).

\item A list of alternative names for the galaxy culled from LEDA.

\item General information about the object, collected from the LEDA data base: 
morphological T-type, morphological type, redshift, and logarithmic size parameters d$_{25}$
and r$_{25}$.

\item Information about the two IUE spectra (SW and LW) presented on the
page: image number, date when each spectrum was acquired, exposure of each 
spectrum in seconds, and position angle of the aperture's major axis.

\item Image numbers of at most ten additional IUE spectra of the object. These 
are selected by decreasing order of exposure time.

\item A plot of the combined UV spectrum of the galaxy, from 1150\AA\, to 3350\AA\,.
Note that the LW spectrum was, in some instances, scaled to match the long wavelength
end of the SW spectrum, as explained below.

\item Two images of the galaxy, extracted from the Digitized Sky Survey (DSS), with the
overlaid outline of the IUE aperture relevant for the specific spectrum displayed here. The
left image is always that of the SW aperture position and the right image is that of the
LW.

\end{enumerate}

The selection of spectra to be combined into the representative UV spectrum of the 
galaxy was done according to the following rules:

\begin{enumerate}

\item Two spectra, one SW and another LW, with the longest exposures, were selected from INES.

\item The positions of the two IUE entrance apertures, for the SW and LW spectra, were
checked in the images with the aperture overlays. If the locations were very similar, the two
spectra were retained and combined into the final representative spectrum. If not, INES was
searched for a suitable pair of deep spectra conforming to the similar-location criterion.

\item In a number of instances, only SW or only LW spectra were available. In these cases,
the left or the right image of the galaxy, with the suitable aperture overlaid, is missing.

\end{enumerate}

The two selected SW and LW spectra were then combined into a single representative
UV spectrum of the object. Sometimes, the mean spectral energy density (SED) levels 
of the SW and LW spectra of the same object were very different from each other. A
simple combination, {\it e.g.} by concatenation of the SW and LW spectra would then
result in a step-like SED. This is clearly visible in some of the spectra displayed in
the ULDA Access Guide No. 3 (Longo \& Capaccioli 1992), {\it e.g.} A1223+4846.

To prevent such an occurence, we decided to bring the LW spectrum to the level of the 
SW so as to ensure a smooth linkage between the spectra. We averaged the flux
density in a 50\AA\, segment at the long wavelength end of the SW spectrum, and 
in a 50\AA\, segment at the short wavelength end of the 
LW spectrum. The normalization constant was determined from these flux
density averages. The difference between the two averages was {\bf added} to the
LW spectrum to bring it into smooth continuation of the SW spectrum, and is indicated
on the spectral plot. This matching
was not performed in cases when it would have driven any part of the LW spectrum to negative 
flux levels. Such cases can be noticed by the gap between the SW and LW spectra displayed 
on the relevant pages ({\it e.g.} IC1613). A marginal case, where the shift was from a
spectrum at zero LW flux level, was for AOO ANON1244-53. In few a cases ({\it e.g.} NGC 1147)
the LW spectrum seems contaminated by additional light between 1900 and 2200\AA\,. In
such cases, no flux-matching procedure was applied.

For some extended galaxies, much larger than the IUE entrance apertures,
spectra were obtained at a number of physically different regions in 
the galaxy. When two spectra, one SW and the other LW, could be
identified in the same location for such a region, one page was dedicated to
the separate presentation of this information. Multiple pages are
shown for the following galaxies: NGC 3034, NGC 3690, NGC 4449, NGC 4861, NGC 5236.

For some spectra, the aperture coordinates reported in the image header are
incorrect and point to a region outside the target, while the spectra show a
 significant signal. For these spectra, a new aperture position was computed
(Solano, E., private communication) based on the guide star position used by the
Guest Observer. Table 2 lists the image number and the revised aperture
position.

We emphasize here that the morphological type of an object included in this
compilation is the one listed in LEDA. Perusing the contents of this INES Guide, we found
that some objects with which we are familiar are obviously mis-classified in LEDA.
Such an example is Mrk 49, an E galaxy in LEDA but which is really a compact, starbursting 
dwarf. It is possible that there are more such mis-classifications, but we have not 
attempted to sort them out. Along with the morphological type listed in Table 1, the numerical
T-type associated with a galaxy should also change. We believe this change not to be 
significant when considering the statistical distribution of the normal galaxy population 
displayed here (cf., Figure 3).

\section{Comparison with the ULDA Access Guide No. 3}

The most important difference between the two guides is inherent in our catalog
being a later compilation, including all galaxies observed by IUE that fulfill the
selection criteria. The improved extraction procedures of INES resulted in much less
noise in the spectrum of a galaxy, in comparison to the spectra presented by
Longo \& Capaccioli (1992). Furthermore, the method of combining the SW and LW
spectra of one object is different; we tried to match the blue end of the LW to the
``red'' end of the SW by averaging regions of overlap and adding a constant to the
LW spectrum, thus the spectra we present are, by definition, smoothly joined.

Longo \& Capaccioli (1992) decided to correct the spectra they plotted in the ULDA 
Access Guide No. 3 for Galactic extinction. They used the Burstein \& Heiles (1984)
color excess and  the Seaton (1979) average extinction curve. It is obvious today
that the
Galactic extinction is extremely patchy, even when using the improved extinction from
Schlegel \etal (1998); we decided therefore not to correct the
spectra for galactic extinction.

The present guide contains 274 objects, approximately twice the number of galaxies 
included in the ULDA guide of normal galaxies.
Therefore, it offers a much wider sample of objects for comparison than the 
previous publication. On the other hand, the same problem that plagued the derivation
of template UV spectra of galaxies by Storchi-Bergman \etal (1994) still remains; in many
cases the IUE aperture samples only the innermost regions of galaxies.

\section{The sample of normal galaxies}

\begin{figure}[tbh]
\vspace{14cm}
\includegraphics{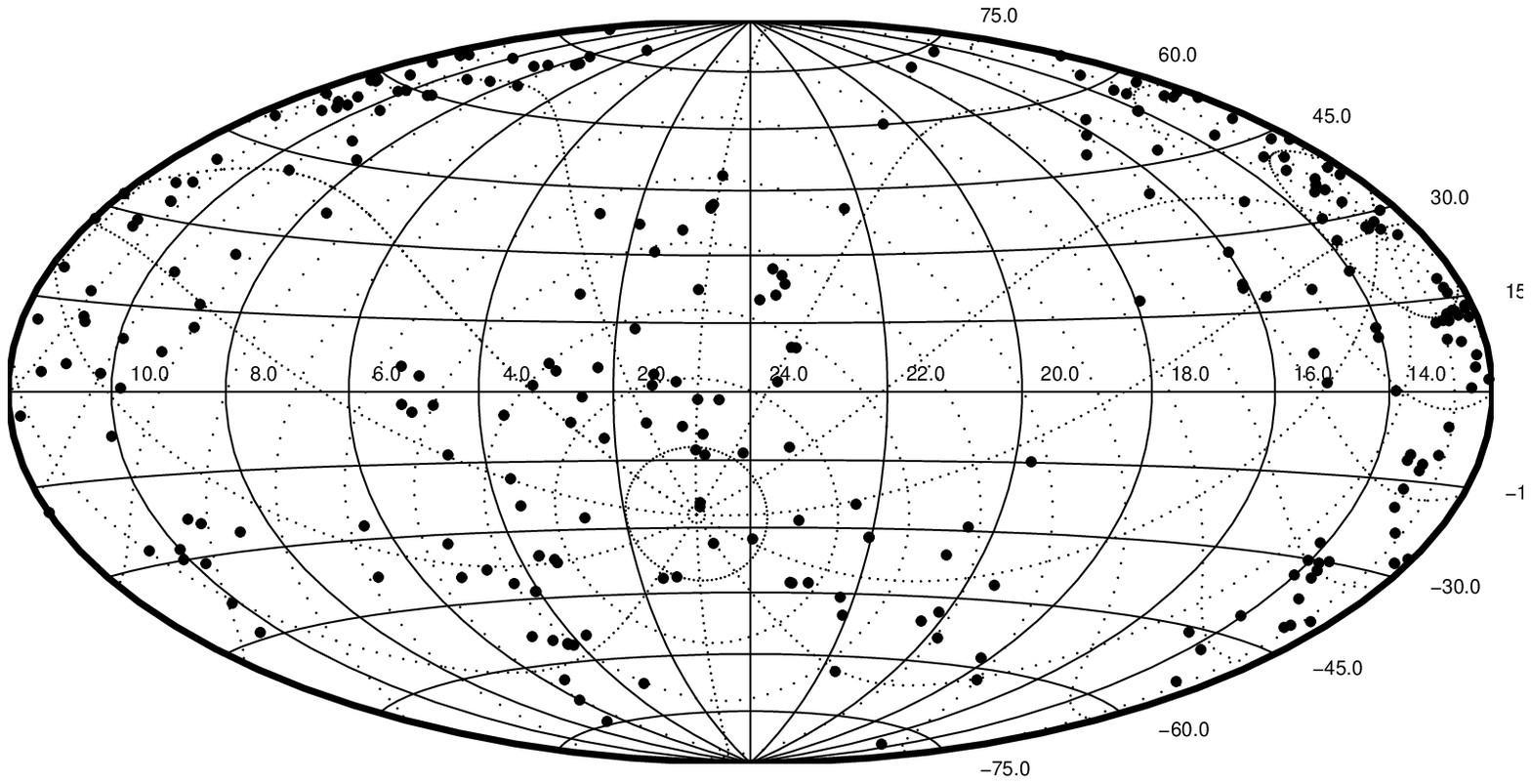}
\caption{Distribution of the normal galaxies included in this Guide, in 
celestial coordinates on an equal-area display. The dotted grid represents 
the Galactic coordinate system. As expected, most UV observations of normal galaxies
concentrate at high Galactic latitudes.}
\end{figure}

The catalog of normal galaxies here presented includes 274 galaxies distributed
over the sky. The distribution is shown in Figure 1.
This large collection of data, resulting from many individual
research projects, does not constitute a homogeneous sample.  
The total B magnitude ${B_{T}}$ (Figure 2) spans a relative wide range (from 6 to 19 mag),
with a mean value of 12.5 and a peak between 13-15 mag, confirming that very few 
faint galaxies were observed by IUE.

\begin{figure}[tbh]
\vspace{14cm}
\includegraphics{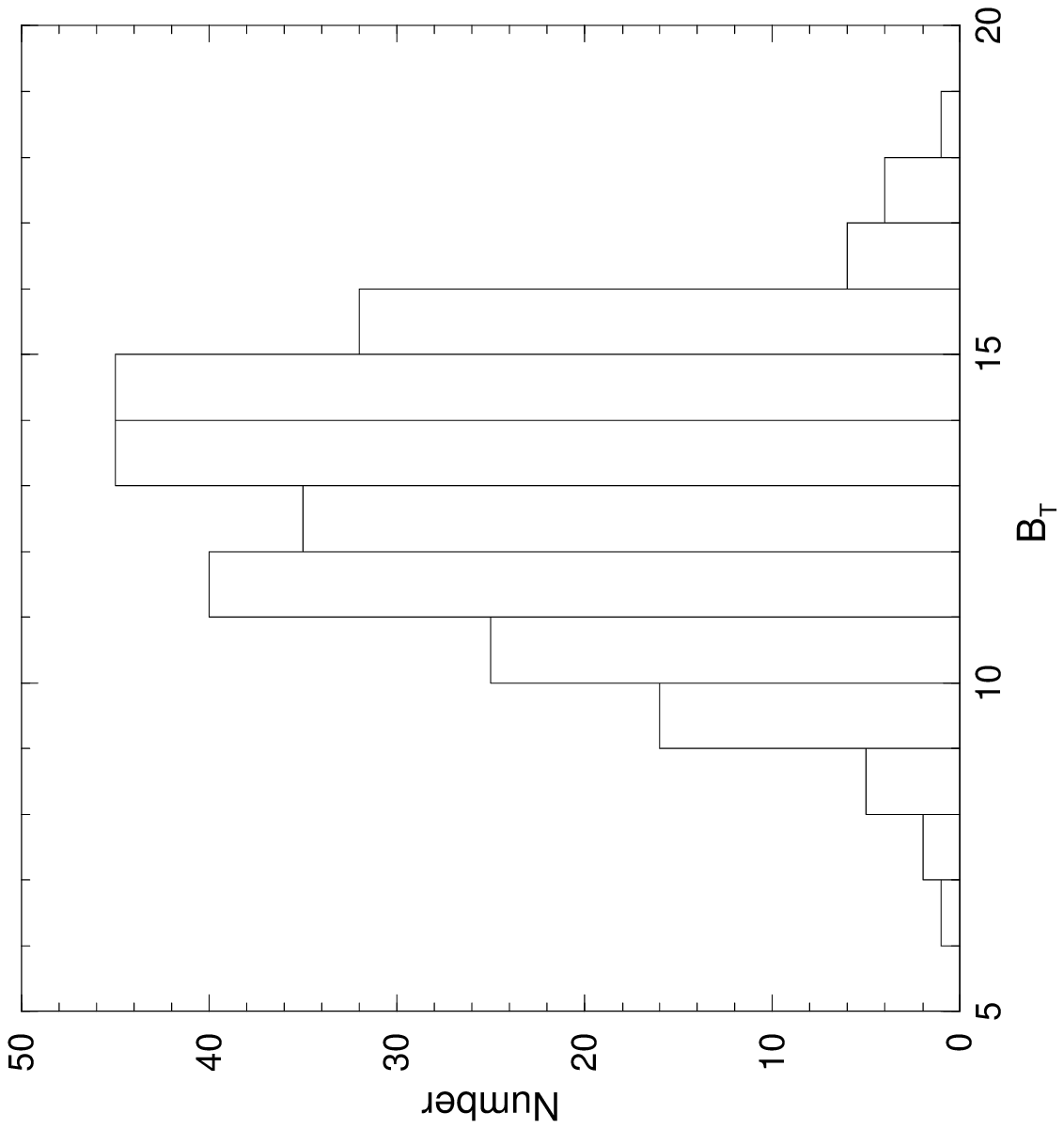}
\caption{Distribution of total B-band magnitudes for the galaxies included in this Guide.}
\end{figure}

The sample contains galaxies of all the morphological types, from  ellipticals
to  irregulars (Figure 3). The elliptical galaxies, some of which are intrinsically 
strong UV emitters
due to the the UV-upturn phenomenon, represent about one-quarter of the total number
of the galaxies in this sample. Figure 3 shows the distribution of galaxies as a 
function of the continuous T-type morphological parameter. The objects were binned 
according to their
morphological classes: E, E-S0, S0, S0a, Sa, Sab, Sb, Sbc, Sc, Scd,
Sd, Sm, Irr.

\begin{figure}[tbh]
\vspace{14cm}
\includegraphics{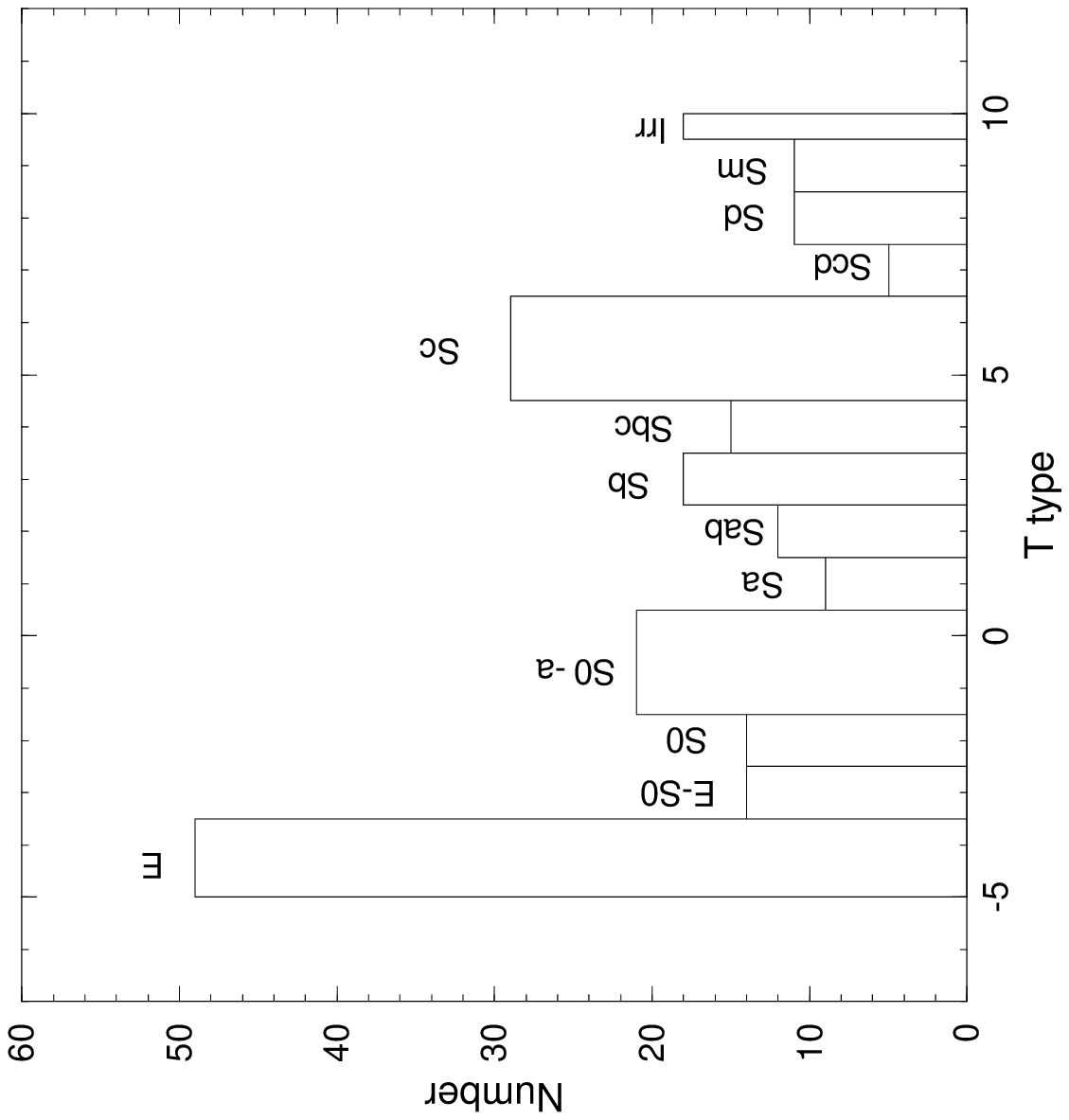}
\caption{Distribution of morphological T-types of galaxies included in this Guide.
The T-types are adopted from LEDA.}
\end{figure} 

The distribution of galaxies with redshift is shown in Figure 4. It is clear that
IUE observed normal galaxies only in the very nearby Universe.

\begin{figure}[tbh]
\vspace{14cm}
\includegraphics{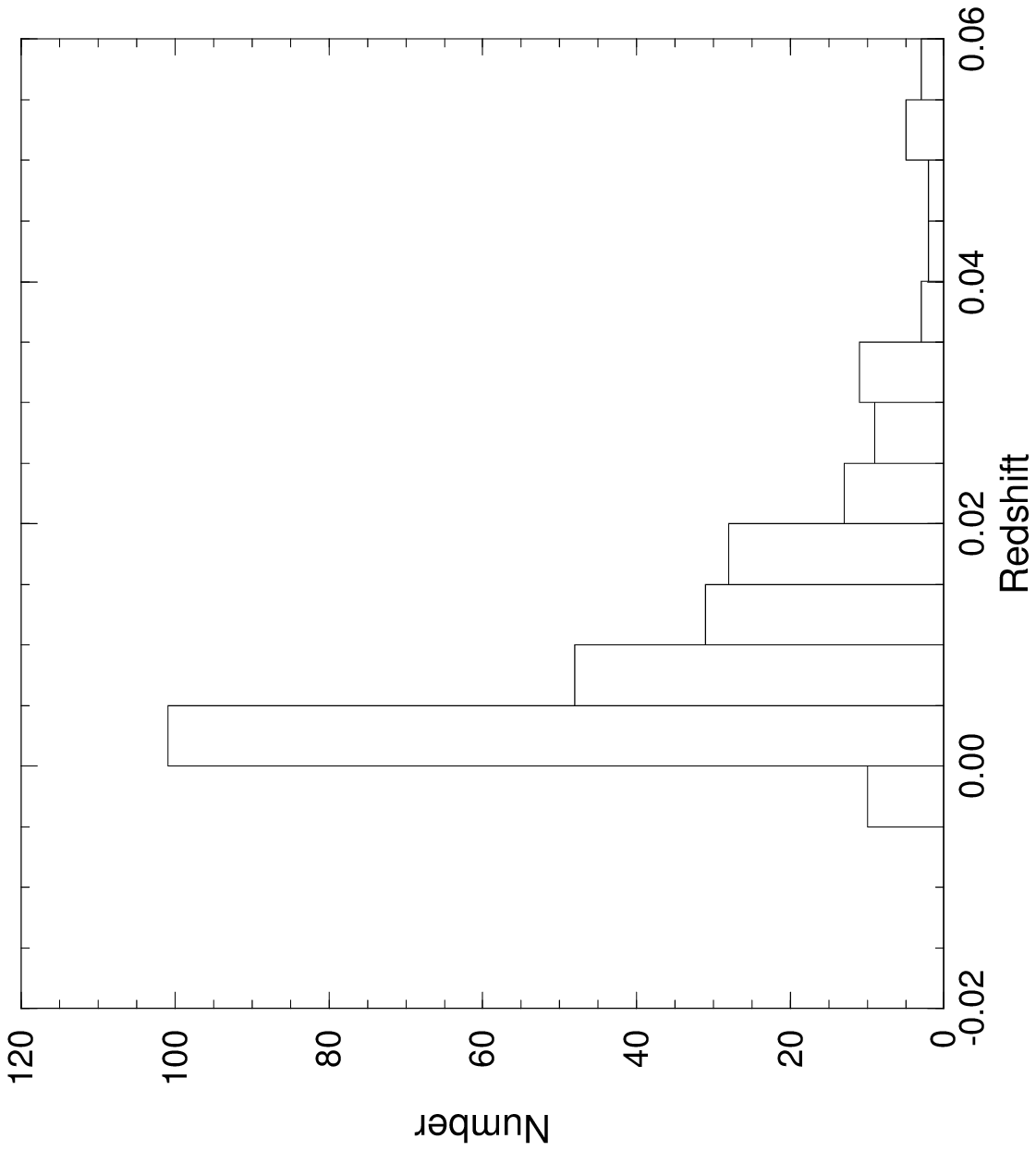}
\caption{Distribution of galaxies included in this Guide vs. redshift.}
\end{figure}

Many of the galaxies observed by IUE are extended objects with respect to
the large aperture of the spectrograh. The large IUE apertures are 
10"$\times$20" and oval-like, each corresponding to the area of a  circular
diaphragm having a diameter of 15.1 arcsec (Longo \& Capaccioli 1992).
In order to estimate the fraction of the galaxy area observed by IUE,
we calculated for each galaxy the ``coverage parameter'' C, defined  as the 
logarithmic  ratio between the surface area of the galaxy and the area of the
large IUE aperture.

\begin{equation}
     C=log[\pi \times (D_{25}^{2} \times R_{25}) / (15.1)^{2} \pi/4] 
\end{equation}

Here D$_{25}$ and R$_{25}$ are the major axis and the axial ratio of the optical 
image of the galaxy as listed in LEDA, in units of arcsec. The numerator
approximates the surface area of the galaxy, represented as an ellipse with 
 the major and minor axes of the galaxy.




A C-value of zero implies that the entire galaxy was measured by in the IUE spectrum.
The distribution of  galaxies as a function of the C parameter is shown 
in Figure 5 . 
For 17 faint galaxies, where the axes are not measurable and which are not shown in 
Figure 5, the C parameter is $ \leq 0$. For $90 \%$  of the sample,
 the IUE aperture covered less than the $10 \% $ of the galaxy.


\begin{figure}[tbh]
\vspace{14cm}
\includegraphics{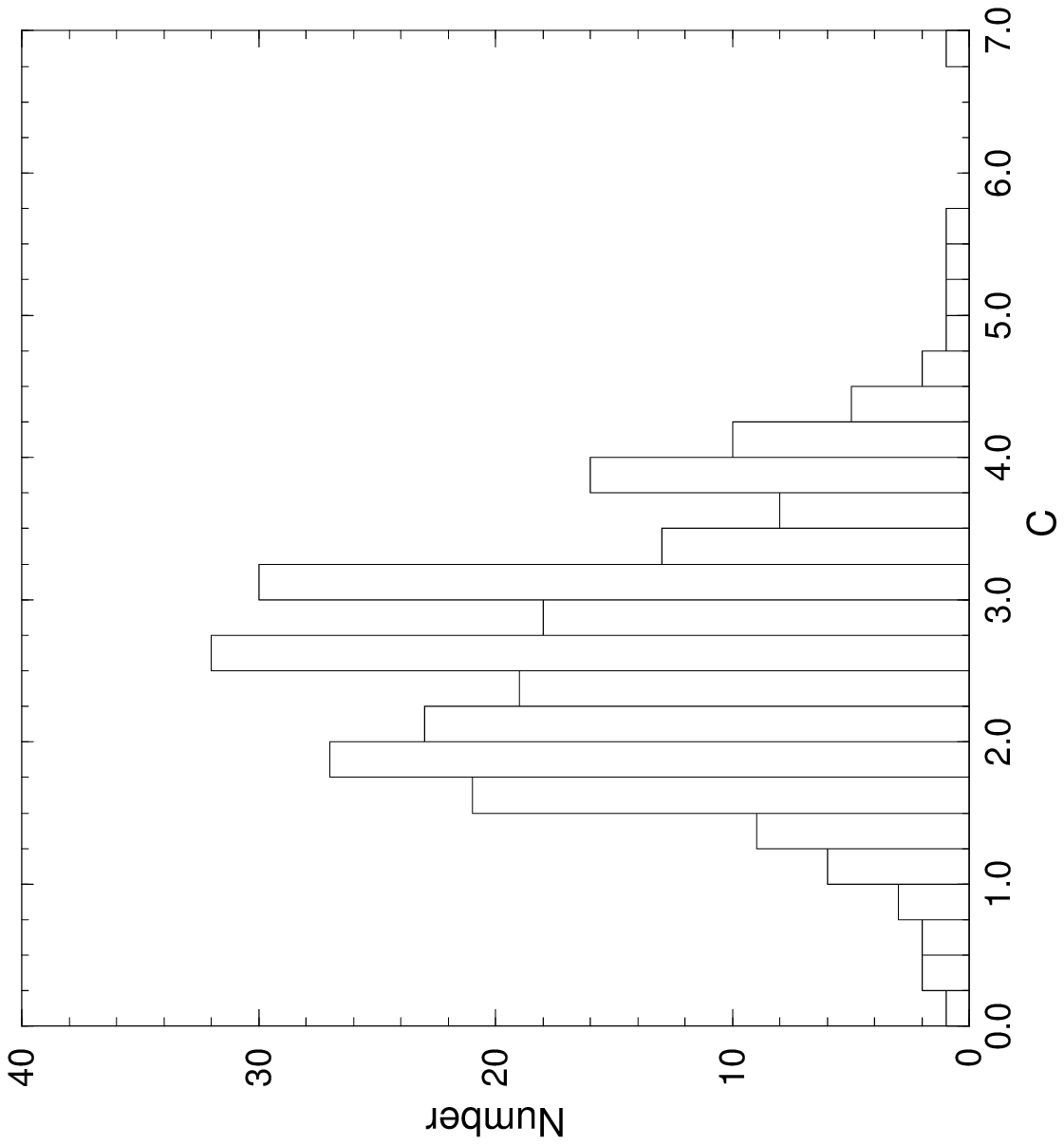}
\caption{Distribution of the logarithmic coverage factor for galaxies included in this Guide.
A value of unity implies that only 10\% of the galaxy was actually measured by IUE. In
most cases, IUE sampled a region that was 1\% to 0.1\% of the target galaxy area.}
\end{figure}

Figure 6 shows the number of available spectra per galaxy. For 
40\% of the galaxies included in this Guide to Normal Galaxies only one spectrum, 
either SW or LW, was obtained. A few galaxies were observed extensively. Among those are
the starburst galaxies NGC 4449, NGC 5253, and NGC 5236.

\begin{figure}[tbh]
\vspace{14cm}
\includegraphics{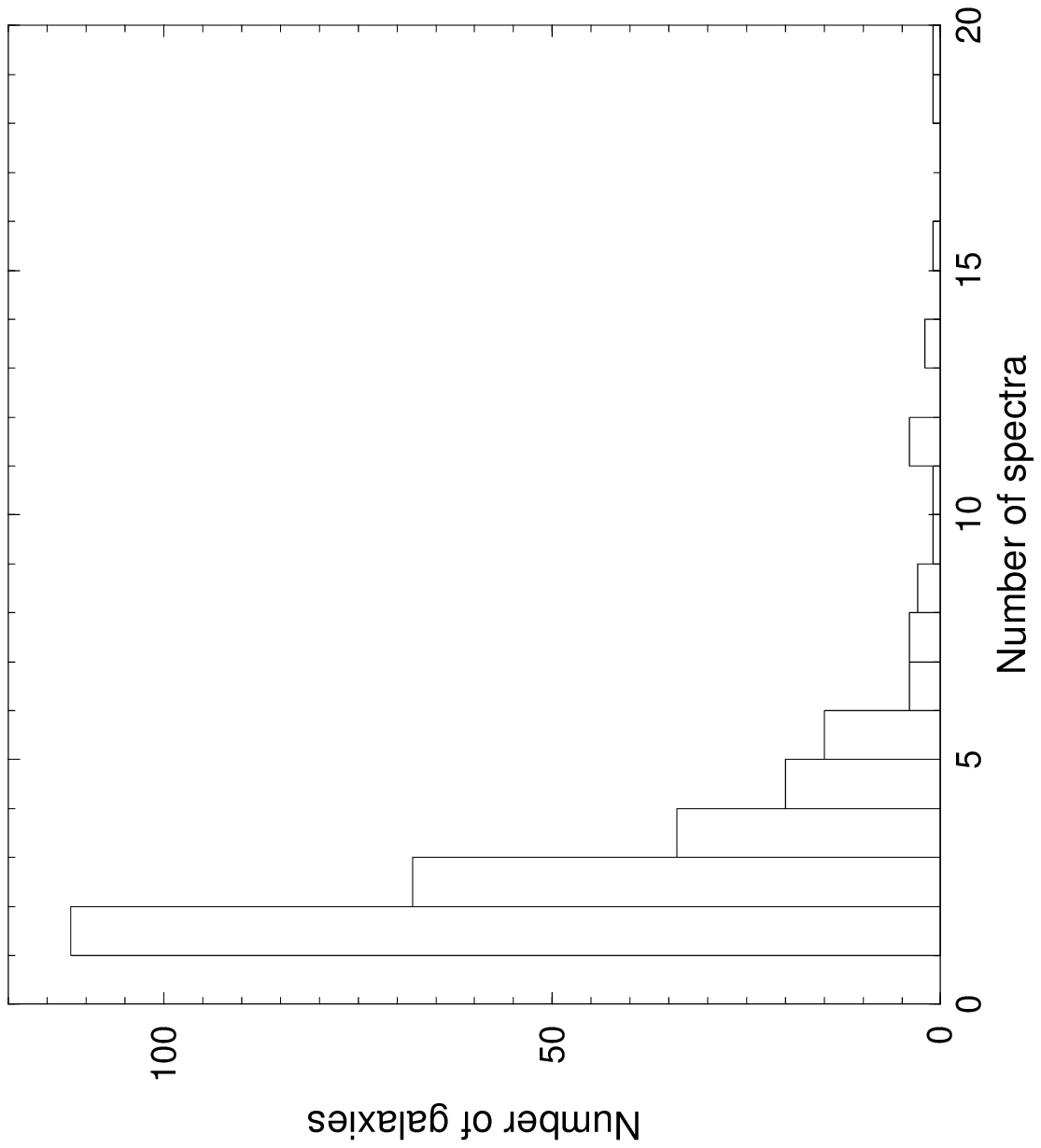}
\caption{Number of spectra available in INES for each galaxy included in this Guide.}
\end{figure}

Figure 7 shows the number of galaxies in our sample  as a function of the
observation date of the two spectra included here. The distribution
indicates two periods of enhanced interest in UV spectra of galaxies. From the launch of
IUE to the mid-80s there was presumably a learning period of UV properties of
galaxies, when many objects were observed. After this, the 
exploitation of IUE began in earnest. The peak in 1994-5 was probably due to ``last-minute''
observations, prior to the closing of the IUE Observatory.

\begin{figure}[tbh]
\vspace{14cm}
\includegraphics{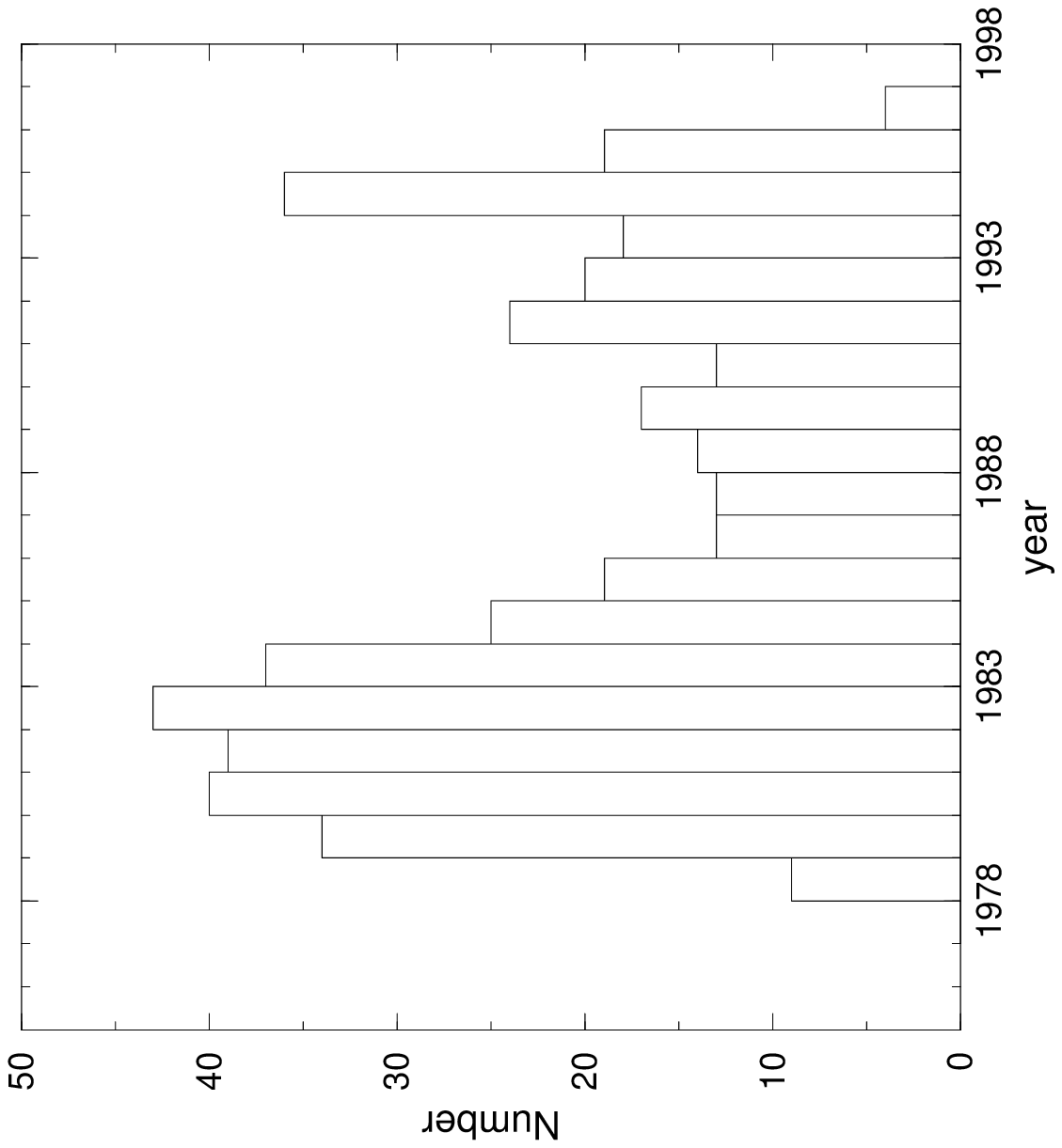}
\caption{IUE observations of normal galaxies vs. the years of observation.}
\end{figure}

In conclusion, the IUE data set for normal galaxies represents mostly a sample
of UV spectra of the innermost regions of  bright nearby galaxies.

\section{Comments of specific objects}

In some cases, the coordinates listed for the entrance aperture do not correspond 
to the coordinates 
of the galaxy identified in the header as the target of the observation, and  
IUE did not, in fact, observe an object but rather sampled blank parts of the sky. 
The galaxies for which the spectra are clearly  
misplaced, and which were removed from the present catalog, are: MCG +06-20-0022, 
NGC 3077, NGC 5813, AOO 1204-137, AOO PKS 1543+091, AOO ACG 1116+51, and AOO POX 4. 
For all the galaxies of the sample but six, the representative spectrum
presented was obtained by the matching procedure of the SW and LW spectra
(see section 5). The additive
constant eventually applied to the LW spectrum is included on the right side of
the plot. The LW spectra of five galaxies: NGC 147, NGC 253, AOO GH 10-4,
NGC 3642, NGC 6062, shows a rising trend toward short wavelengths; 
it is possible that  an additional light source contributed to the recorded spectrum. 
The LW spectrum for these galaxies is presented   
without performing the matching  procedure.


Below we provide a few  comments on individual objects and for the
galaxies for which more than one page is presented:

AOO UM 239: This object was rejected from the list of galaxies included in this
guide. The IUE aperture is located on an object that is one arcmin 
        south of the declination quoted by NED for AOO UM 239.

NGC 221: This companion of the Andromeda galaxy, in the Local Group, is one of 
    the most observed elliptical galaxies.
        The CIV line is absent in the deepest image:  SWP30289.

NGC 224: This is M31, a spiral galaxy in the Local Group where the IUE spectra sampled
        the bulge population at the center. The UV spectral energy
        distribution appears similar to that of an elliptical galaxy. 

NGC 244: The single spectrum of this galaxy available in the archive, belongs
actually to the galaxy 
        NGC 224, according to the aperture coordinates.

IC 1613: The SW and LW images refer to different regions
        of the galaxy. Therefore, the corresponding spectra displayed in 
        the figure are separated. 

NGC 300: This very extended galaxy was removed from the list. Only
        one IUE spectrum was obtained for it, and it corresponds to a
    region at the
        outskirts of the galaxy, not to a specific significant
        region such as the photocenter. 

NGC 1022: The signal of the SW spectrum is zero. No matching procedure
        was applied to the LW spectrum, for obvious reasons.

NCG 1617: Some pixels of the LW spectrum are saturated.

NGC 1705: Although some pixels of the LW spectrum are saturated, the final
         representative spectrum is reasonably smooth.

PG 0833+652: Some pixels of the LW spectrum are saturated.

MRK 116 A: This object is one of a pair of interconnected blue compact
galaxies.
        As a low metallicity galaxy, it is a good target for studying 
        the star formation history and chemical mixing process.

NGC 3034: This nuclear starburst galaxy, also known as M82, shows a bipolar
        outflow along its minor axis. The IUE aperture was  located
        on its center and on two regions near the outflow, named 
        BD+ 70 5888A KNOT A and  BD+ 70 5888A KNOT B.
        This galaxy has been extensively studied at various wavelengths.
       
AOO GH 10-4: The coordinates listed for this galaxy are from NED.

AOO ACG 1116+51: The IUE coordinates of this target are wrong. The IUE
        spectrograph missed the object, which is at 11$^h$ 19$^m$ 34$^s$.24 
+51$^{\circ}$ 30' 11".84
        (PMM USNO-A2.0). As explained above, this galaxy was excluded from 
our compilation.

AOO 1214-277 and  AOO 1214-28: These targets are near the Tololo 21
        galaxy. The coordinates reported in the corresponding pages 
        are those of some objects near the galaxy, listed in the PMM USNO-A2.0
catalog.

NGC 4449: Up to 48 spectra were obtained by IUE for NGC 4449.
         There are many HII regions and a complex of supernova remnants in this 
         irregular galaxy. Eight pages are shown here for NGC 4449, representing
         the different regions sampled by IUE.
          
AOO ANON 1244-53: The level of the spectrum in the LW frame is consistent with a noisy 
         zero signal.

NGC 4861: This system consists of IC 3961 plus a blue compact HII region in
         the south-west, identified as NGC 4861. Individual pages are shown
         for the two objects.
          
AOO POX 120: The declination quoted in LEDA is wrong. The coordinates listed here
       for this galaxy are from NED.

NGC 5236: This galaxy, better known as M83, is a nearby bright starburst spiral.
        Young massive star cluster systems  are located in the nucleus and in
        the arms. Many  IUE spectra were obtained at the position of some
        of the large number of historical supernovae  observed in this galaxy. 

NGC 5253: This nearby dwarf starburst galaxy hosts many bright knots embedded 
       in its amorphous central region. It is believed that this complex of
       starburst regions is the source of the strong soft thermal X-ray 
       emission detected by ROSAT. The representative spectrum is quite smooth, 
       although the SW spectrum  is saturated from 1800 to 1900\AA.

NGC 5461 and NGC 5471: These are two HII regions in the outskirts of M101.

AOO PKS 1543+091: The IUE coordinates are wrong, and the spectrograph
         pointed at an unidentified faint object. The object is not included in
our ``normal galaxies'' sample.

ESO B338-IG4: This is a compact galaxy with double nuclei that was extensively 
studied by IUE.
             
AOO AM 2020-504; The declination listed in LEDA is wrong. The coordinates presented
       here for this galaxy are from NED.


MCG-07-47-023: The signal in the SW and LW spectra is consistent with 
a noisy zero signal.

NGC 7673: The coordinates of the LW frame are outside the object. 
        The correct location of the aperture, calculated from the
        position of the guide star as stored in the header of this
        image, is still outside the galaxy.

\section{Conclusions}

We present here a compilation of all galaxies classified as ``normal'', {\it i.e.} not
containing an AGN, which were observed by the IUE satellite. This is an updated and
enlarged version of the previous IUE ULDA Guide to Normal Galaxies, produced
by Longo \& Capaccioli, and supersedes it. We took advantage of the final 
processing of IUE observations
within the INES archive to extract all the observations of galaxies classified as 
normal and we  present their UV spectra after matching the fluxes between the SW
and LW segments that were observed toward the same location in the galaxy.

We believe that although most IUE observations of normal galaxies were only of 
their central regions, as defined by the optical light distributions, this compilation
will be of some use in understanding the UV light produced by the different
stellar populations in objects lacking a central, compact, massive object.

\section*{Acknowledgments}

This compilation was produced in support of the TAUVEX mission.
    UV research at Tel Aviv University is supported by grants from
    the Ministry of Science, Culture and Sport through the Israel Space Agency,
    from the Austrian Friends of Tel Aviv
    University, and from a Center of Excellence Award from the Israel
    Science Foundation. NB acknowledges support from an US-Israel Binational
    Award to study UV sources measured by the FAUST experiment. In this
publication we made use of the LEDA database, http://www-obs.univ-lyon1.fr.

We are grateful to Willem Wamsteker for encouraging us to go ahead with the
production of this Guide. Enrique Solano helped us by recalculating the true
pointing of IUE of some 20 discrepant IUE spectra. This solved the 
mystery of reasonable spectra apparently obtained on blank sky.

\newpage

\section*{References}
\begin{description}

 
   
 

    \item Bianchi, L. \& Martin, C. 1997, in ``The Ultraviolet 
    Astrophysics beyond
    the IUE final Archive'' (R. Harris, ed.) ESA SP-413, 797.

\item Bonatto, C., Bica, E., Pastoriza, M.G. \& Alloin, D. 1996,
Astron. Astrophys. Suppl. {\bf 118}, 89.
  
    \item Bowyer, S., Sasseen, T., Wu, X. \& Lampton, M. 1995, 
    Astrophys. J. Suppl {\bf 96}, 461. 

\item Boyarchuk, A.A. 1994, ``Astronomical investigations with the ASTRON cosmic
station'' (in Russian) Moscow: FIZMATLIT.

    \item Brosch, N. 1991, Mon. Not. R. astr. Soc. {\bf 250}, 780. 

    \item Brosch, N., Almoznino, E., Leibowitz, E.M., Netzer, H., 
    Sasseen, T.P., Bowyer, S., Lampton, M. \& Wu, X. 1995, Astrophys. J. {\bf 
    450}, 137. 
 
    \item Brosch, N., Formiggini, L., Almoznino, E., Sasseen, T., 
    Lampton, M. \& Bowyer, S. 1997, Astrophys. J. Suppl. {\bf 111}, 143.  
 
    \item Brosch, N., Ofek, E., Almoznino, E., Sasseen, T., Lampton, M. 
    \& Bowyer, S. 1998, Mon. Not. R. astr. Soc., {\bf 295}, 959.

\item Brosch, N., Shara, M., MacKenty, J., Zurek, D. \& McLean, B. 1999,
Astron. J, {\bf 117}, 206.


    \item Burstein, D., Bertola, F., Buson, L.M., Faber, S.M. \& Lauer, 
    T.R. 1988, Astrophys. J. {\bf 328}, 440. 

\item Burstein, D. \& Heiles, C. 1984, ApJS, {\bf 54}, 33.

    \item  Courvoisier, T.J.-L., Paltini, S. 1992, ``IUE ULDA Access Guide No. 4:
     Active Galactic Nuclei, ESA SP-1153 (Vol. A \& B)

    \item Deharveng, J.-M., Sasseen, T.P., Buat, V., Bowyer, S., 
    Lampton, M. \& Wu, X. 1994, Astron. Astrophys. {\bf 289}, 715. 

    \item Donas, J., Deharveng, J.-M., Laget, M., Milliard, B. \& 
    Huguenin, D. 1987, Astron. Astrophys. {\bf 180}, 12.  

    \item Ellis, R.S., Gondhalekar, P.M. \& Efstathiou, G.
1982, Mon. Not. R. astr. Soc. {\bf 201}, 223.
 
    \item Ellis, R.S. 1997, Ann. Rev. Astron. Astrophys. {\bf 35}, 389.

    \item Fanelli, M.N., O'Connell, R.W. \& Thuan, T.X. 1987, Astrophys. 
    J. {\bf 321}, 768.

    \item Heck, A., Egret, D., Jaschek, M. \& Jaschek, C. 1984, 
    ``IUE Low-Dispersion Spectra Reference Atlas - Part 1. Normal Stars'',
    Paris: ESA SP-1052.

    \item Kinney, A.L., Bohlin, R.C., Calzetti, D., Panagia, N. \& 
    Wyse, R.F.G. 1993, Astrophys. J. Suppl. {\bf 86}, 5. 

\item Kinney, A.L., Calzetti, D., Bohlin, R.C., McQuade, K., Storchi-Bergmann, T.
\& Schmitt, H.R. 1996, Astrophys. J., {\bf 467}, 38.

    \item Longo, G., Capaccioli, M.\& Ceriello, A. 1991, Astron. 
    Astrophys. Suppl. {\bf 90}, 375.

    \item Longo, G. \& Capaccioli, M. 1992, ``IUE ULDA Access Guide No. 3:
    Normal Galaxies'', ESA SP-1152.

    \item Milliard, B., Donas, J., Laget, M., Armand, C. \& Vuillemin, 
    A. 1992, Astron. Astrophys. {\bf 257}, 24.

    \item McQuade, K., Calzetti, D. \& Kinney, A.L. 1995, Astrophys. J. 
    Suppl., {\bf 97}, 331. 
 
\item Nichols, J.S., Garhart, M.P., De La Pena, M.D. \& Levay, K.L. 1993, "IUE NEWSIPS
Image Processing System Information Manual: Low Dispersion Data", Version
1.0, CSC/SD-93/6062.

    \item O'Connell, R.W. 1992, in ``The stellar population of galaxies''
    (B. Barbuy and A. Renzini, eds.) Dordrecht: Reidel Publishing Co., p. 233. 
 
    \item O'Connell, R.W., Bohlin, R.C., Collins, N.R.,
     Cornett, R.H., Hill, J.K., Hill, R.S., Landsman, W.B.,
    Roberts, M.S., Smith, A.M. \& Stecher, T.P. 1992, Astrophys. 
    J. Lett. {\bf  395}, L45. 

\item O'Connell, R.W. 1999, Ann. Rev. Astron. Astrophys., {\bf 37}, 603.

   \item Rifatto, A., Longo, G. \& Capaccioli, M. 1995a, Astron. 
    Astrophys. Suppl. {\bf 109}, 341. 
 
    \item Rifatto, A., Longo, G. \& Capaccioli, M. 1995b, Astron. 
    Astrophys. Suppl. {\bf 114}, 527. 

\item Rodriguez-Pascual, P.M., Schartel, N., Wamsteker, W. \& Perez-Calpena, A.
1998, "Extraction of Low Dispersion Spectra for INES", INES Document 4.3.

\item Schartel, N. \& Rodriguez-Pascual, P.M. 1998, "INES Noise Model", INES
Document 4.2.

\item Schartel, N. \& Skillen, I. 1998, in "Ultraviolet Astrophysics beyond the IUE 
Final Archive", ESA SP-413 (W. Wamsteker \& R. G0nzales Riestra, eds.), p. 735.

\item Schlegel, D.J., Finkbeiner, D.P. \& Davis, M. 1998, ApJ, {\bf 500}, 525.

\item Seaton, M.J. 1979, Mon. Not. R. astr. Soc. {\bf 187}, 73.

    \item Storchi-Bergmann, T., Calzetti, D. \& Kinney, A.L. 1994, 
    Astrophys. J. {\bf 429}, 572  (SB2). 

\item Steindling, S., Brosch, N. \& Rakos, K. 2000, Astrophys. J., submitted.

     \item Wu, C.-C., Crenshaw, D.M., Blackwell, J.H., Wilson-Diaz, D.
     Schiffer, F.H., Burstein, D., Fanelli, M.N. \& O'Connell, R.W. 1991, ``IUE
     Ultraviolet Spectral Atlas'', IUE NASA Newsletter No. 43.

\end{description}

\newpage

\begin{deluxetable}{cccccrccccrccr}
\tablecaption{Galaxies included in this publication}
\small
\tablehead{\colhead{IUE ID} & \colhead{N$_{sp}$} & \colhead{h } & \colhead{m} & \colhead{s} & 
\colhead{$^{\circ}$} & \colhead{'} & \colhead{"}  & \colhead{Morph. Type} &
\colhead{T} & \colhead{B$_T$} & \colhead{B--V} & \colhead{U--B} }
\startdata
 NGC7828         & 1 &00 &06 &27 &  -13 &24 &54 &Sc       &  5.484 &  14.37&   0.51&  -0.20  \nl  
 NGC118          & 1 &00 &27 &16 &  -01 &46 &48 &E        & -5.000 &  14.63&       &         \nl  
 NGC147          & 2 &00 &33 &12 &   48 &30 &28 &E        & -4.785 &  10.37&   0.95&         \nl  
 ESOB350-IG38    & 1 &00 &36 &53 &  -33 &33 &24 &S?       & 10.000 &  15.45&       &         \nl  
 NGC205          & 9 &00 &40 &23 &   41 &41 &11 &E        & -4.763 &   8.73&   0.86&         \nl
 NGC210          & 2 &00 &40 &35 &  -13 &52 &26 &SBb      &  2.986 &  11.83&   0.71&   0.07  \nl    
 ABCG85          & 5 &00 &41 &51 &  -09 &18 &15 &E-S0     & -3.460 &  14.50&   1.00&         \nl  
 NGC221          &10 &00 &42 &42 &   40 &51 &55 &E        & -4.712 &   8.73&   0.95&   0.48  \nl  
 NGC224          &18 &00 &42 &44 &   41 &16 &08 &Sb       &  2.996 &   4.18&   0.93&   0.50  \nl   
 NGC245          & 1 &00 &46 &06 &  -01 &43 &20 &Sb       &  3.129 &  12.82&       &         \nl  
 ESOB474- 26     & 1 &00 &47 &08 &  -24 &22 &14 &Sc       &  4.826 &  14.94&       &         \nl  
 NGC253          & 4 &00 &47 &33 &  -25 &17 &18 &SBc      &  5.055 &   7.92&   0.86&   0.38  \nl  
 IC1586          & 1 &00 &47 &56 &   22 &22 &28 &compact  &        &  14.90&       &         \nl
 MRK960          & 2 &00 &48 &35 &  -12 &43 &01 &S0       & -1.984 &  13.97&       &         \nl  
 NGC337          & 1 &00 &59 &50 &  -07 &34 &41 &SBcd     &  6.905 &  12.03&   0.45&  -0.09  \nl  
 IC1613          & 5 &01 &04 &54 &   02 &07 &60 &Irr      &  9.870 &   9.94&   0.68&         \nl  
 NGC404          & 5 &01 &09 &27 &   35 &43 &04 &E-S0     & -2.798 &  10.97&   0.94&   0.28  \nl  
 ESOB296-IG11    & 4 &01 &19 &57 &  -41 &14 &10 &Sd       &  7.901 &  14.47&   0.26&   0.34  \nl  
 NGC520          & 1 &01 &24 &35 &   03 &47 &25 & S?      &  3.840 &  12.14&   0.82&   0.17  \nl  
 AOO PKS0123-16  & 5 &01 &25 &48 &   01 &22 &18 &Irr      & 10.000 &  14.42&       &         \nl  
 NGC584          & 1 &01 &31 &21 &  -06 &52 &06 &E        & -4.620 &  11.31&   0.96&   0.50  \nl  
 NGC598          &13 &01 &33 &51 &   30 &39 &37 &Sc       &  5.961 &   6.20&   0.56&  -0.09  \nl  
 NGC625          & 2 &01 &35 &05 &  -41 &26 &11 &SBm      &  9.295 &  11.60&   0.56&  -0.06  \nl  
 NGC660          & 1 &01 &43 &01 &   13 &38 &37 &SBa      &  1.091 &  11.83&   0.87&         \nl  
 MRK2            & 1 &01 &54 &53 &   36 &55 &02 &SBa      &  0.599 &  13.92&   0.59&  -0.14  \nl  
 NGC835          & 1 &02 &09 &25 &  -10 &08 &09 &SBab     &  1.945 &  12.97&   0.81&   0.19  \nl
 IC214           & 1 &02 &14 &06 &   05 &10 &32 &Sd       &  7.500 &  14.68&   0.53&  -0.27  \nl  
 NGC936          & 1 &02 &27 &38 &  -01 &09 &17 &S0-a     & -0.983 &  11.01&   0.98&   0.56  \nl  
 NGC992          & 1 &02 &37 &26 &   21 &05 &55 &Sc       &  5.116 &  15.56&   0.41&  -0.43  \nl  
 NGC1022         & 2 &02 &38 &33 &  -06 &40 &41 &SBab     &  1.867 &  12.09&   0.75&   0.24  \nl  
 IC1830          & 1 &02 &39 &04 &  -27 &26 &43 &S0-a     & -0.862 &  13.34&   0.41&  -0.25  \nl  
 NGC1023         & 1 &02 &40 &24 &   39 &03 &46 &E-S0     & -2.610 &  10.12&   1.00&   0.56  \nl  
 MRK600          & 1 &02 &51 &04 &   04 &27 &09 &Sc       &  5.000 &  14.99&       &         \nl  
 ABCG400A        & 1 &02 &57 &42 &   06 &01 &38 &E        & -4.250 &  13.86&   1.16&   0.67  \nl  
 IC298A          & 1 &03 &11 &20 &   01 &18 &46 &SBbc     &  3.640 &  15.98&       &         \nl  
 NGC1313         & 4 &03 &18 &15 &  -66 &29 &51 &SBcd     &  6.990 &   9.66&   0.50&  -0.23  \nl  
 NGC1316         & 2 &03 &22 &42 &  -37 &12 &28 &S0       & -1.738 &   9.37&   0.89&   0.39  \nl  
 NGC1326         & 2 &03 &23 &56 &  -36 &27 &50 &S0-a     & -0.776 &  11.39&   0.88&   0.29  \nl  
 NGC1385         & 2 &03 &37 &28 &  -24 &30 &12 &SBc      &  6.004 &  11.45&   0.51&  -0.17  \nl  
AOO SBSG 0335-052 & 3 &03 &37 &47 &  -05 &02 &47 &EmLS     &        &       &       &         \nl  
 NGC1399         & 3 &03 &38 &29 &  -35 &26 &58 &E        & -4.548 &  10.39&   0.96&   0.50  \nl  
 NGC1404         & 4 &03 &38 &52 &  -35 &35 &36 &E        & -4.749 &  10.89&   0.98&   0.56  \nl  
 NGC1407         & 4 &03 &40 &12 &  -18 &34 &52 &E        & -4.550 &  10.69&   1.03&         \nl  
 ESOB156-IG7     & 3 &03 &41 &12 &  -54 &00 &39 &S0       &  2.000 &  15.71&       &         \nl  
 NGC1510         & 3 &04 &03 &33 &  -43 &24 &01 &S0       & -1.949 &  13.51&   0.45&  -0.19  \nl  
 NGC1512         & 4 &04 &03 &55 &  -43 &21 &03 &SBa      &  1.138 &  11.05&   0.81&   0.17  \nl  
 NGC1533         & 1 &04 &09 &51 &  -56 &07 &14 &S0       & -2.396 &  11.82&   0.99&   0.50  \nl  
 NGC1549         & 2 &04 &15 &45 &  -55 &35 &31 &E        & -4.283 &  10.69&   0.94&   0.50  \nl  
 NGC1553         & 3 &04 &16 &10 &  -55 &46 &51 &S0       & -2.344 &  10.31&   0.88&   0.48  \nl  
AOO TOL 0420-414 & 1 &04 &21 &59 &  -41 &19 &21 &EmLS     &        &  18.26&       &         \nl  
 NGC1617         & 1 &04 &31 &39 &  -54 &36 &05 &S0-a     &  0.183 &  11.29&   0.94&   0.50  \nl
 ABCG496         & 2 &04 &33 &38 &  -13 &15 &38 &E        & -4.067 &  13.83&       &         \nl  
 NGC1637         & 2 &04 &41 &28 &  -02 &51 &30 &SBc      &  5.029 &  11.38&   0.64&   0.05  \nl  
AOO TOL 0440-381 & 1 &04 &42 &08 &  -38 &01 &03 &EmLS     &        &       &       &         \nl
 NGC1705         &11 &04 &54 &14 &  -53 &21 &44 &E-S0     & -2.963 &  12.77&   0.38&  -0.44  \nl  
 NGC1691         & 1 &04 &54 &38 &   03 &16 &06 &S0-a     &  0.064 &  12.77&       &         \nl  
 NGC1741         & 5 &05 &01 &38 &  -04 &15 &25 &Sm       &  8.913 &  15.52&       &         \nl  
 NGC1800         & 1 &05 &06 &26 &  -31 &57 &16 &Irr      &  9.908 &  13.04&   0.55&  -0.17  \nl  
 MRK1094         & 1 &05 &10 &48 &  -02 &40 &54 &S?       &  3.400 &  13.87&       &         \nl  
 NGC1819         & 1 &05 &11 &46 &   05 &12 &01 &S0       & -1.972 &  13.38&       &         \nl  
AOO TOL 0513-393 & 1 &05 &15 &20 &  -39 &17 &41 &EmLS     &        &       &       &         \nl  
 NGC1947         & 1 &05 &26 &48 &  -63 &45 &41 &E-S0     & -3.165 &  11.70&   1.01&   0.50  \nl  
AOO GH 10- 4     & 2 &06 &00 &46 &  -68 &40 &00 &SBb      &        &       &       &         \nl  
 NGC2146         & 1 &06 &18 &40 &   78 &21 &19 &SBab     &  2.268 &  11.16&   0.80&   0.29  \nl  
 NGC2217         & 3 &06 &21 &40 &  -27 &14 &00 &S0-a     & -0.639 &  11.63&   1.00&   0.55  \nl  
 AOO 0644-741    & 1 &06 &43 &00 &  -74 &14 &11 & E-S0    & -2.970 &  13.88&   0.88&   0.19  \nl  
AOO TOL 0645-376 & 1 &06 &46 &49 &  -37 &43 &25 &EmLS     &        &  17.38&       &         \nl  
 MRK7            & 2 &07 &28 &11 &   72 &34 &20 &Sd       &  7.752 &  14.44&       &         \nl  
 IC2184          & 2 &07 &29 &25 &   72 &07 &41 &Sbc      &  4.077 &  14.38&       &         \nl  
 NGC2403         & 5 &07 &36 &54 &   65 &35 &58 &SBc      &  5.967 &   8.83&   0.47&         \nl  
 AOO HARO 1      & 2 &07 &36 &57 &   35 &14 &33 &Irr      &  9.711 &  12.72&   0.42&  -0.20  \nl  
 MRK12           & 3 &07 &50 &48 &   74 &21 &32 &SBc      &  4.987 &  13.11&   0.44&  -0.38  \nl  
 NGC2537         & 4 &08 &13 &15 &   45 &59 &29 &SBm      &  8.617 &  12.19&   0.63&  -0.14  \nl  
MCG +12-08-0033  & 2 &08 &19 &06 &   70 &42 &51 &Irr      &  9.813 &  11.10&   0.44&         \nl  
 PK 248+08 1     & 2 &08 &36 &16 &  -26 &24 &40 &E-S0     & -2.621 &  12.47&       &         \nl  
MCG +12-08-0048  & 3 &08 &37 &03 &   69 &46 &29 &Irr      &  9.867 &  15.30&       &         \nl
 PG 0833+652     & 4 &08 &38 &23 &   65 &07 &16 &pec      &        &       &       &         \nl  
 NGC2623         & 3 &08 &38 &24 &   25 &45 &01 &Sb       &  2.700 &  13.99&   0.63&   0.12  \nl  
 AOO T 0840+120  & 1 &08 &42 &21 &   11 &50 &01 &         &        &       &       &         \nl  
 MRK702          & 2 &08 &45 &34 &   16 &05 &48 &compact  &        &  15.57&       &         \nl  
 NGC2684         & 2 &08 &54 &53 &   49 &09 &38 &Sc       &  5.186 &  13.65&   0.69&  -0.20  \nl  
 NGC2773         & 1 &09 &09 &44 &   07 &10 &26 &S?       &  2.700 &  14.46&       &         \nl  
 NGC2768         & 2 &09 &11 &38 &   60 &02 &22 &E        & -4.422 &  10.80&   0.98&   0.47  \nl  
 NGC2784         & 2 &09 &12 &19 &  -24 &10 &22 &S0       & -2.075 &  11.17&   1.14&   0.73  \nl  
 MRK19           & 1 &09 &16 &43 &   59 &46 &20 &BCG      &        &  15.50&       &         \nl  
 NGC2798         & 3 &09 &17 &23 &   42 &00 &02 &SBa      &  1.068 &  13.04&   0.73&  -0.01  \nl  
 IC2458          & 2 &09 &21 &29 &   64 &14 &11 &S0-a     &  0.000 &  15.41&   0.41&  -0.63  \nl  
 NGC2865         & 2 &09 &23 &31 &  -23 &09 &48 &E        & -4.115 &  12.41&   0.92&   0.41  \nl  
 NGC2903         & 5 &09 &32 &10 &   21 &30 &02 &SBbc     &  4.006 &   9.48&   0.68&   0.06  \nl  
 MRK116 A        &11 &09 &34 &02 &   55 &14 &25 &compact  &        &  15.61&   0.11&  -0.64  \nl  
 NGC2997         & 3 &09 &45 &39 &  -31 &11 &28 &SBc      &  5.076 &  10.08&       &         \nl  
 NGC3023         & 2 &09 &49 &53 &   00 &37 &13 &SBc      &  5.473 &  12.95&       &         \nl  
 NGC3049         & 3 &09 &54 &50 &   09 &16 &19 &SBab     &  2.466 &  13.27&       &         \nl  
 NGC3034         & 5 &09 &55 &54 &   69 &40 &57 &Sd       &  7.932 &   9.07&   0.89&   0.31  \nl  
 ESOB435-IG20    & 1 &09 &59 &21 &  -28 &07 &54 &Sb       &  3.000 &  14.40&       &         \nl  
 MRK25           & 3 &10 &03 &52 &   59 &26 &11 &E-S0     & -3.243 &  14.82&   0.57&  -0.27  \nl  
 NGC3115         & 5 &10 &05 &14 &  -07 &43 &07 &E-S0     & -2.807 &   9.86&   0.98&   0.55  \nl  
 NGC3125         & 3 &10 &06 &34 &  -29 &56 &10 &E        & -4.931 &  13.47&   0.50&  -0.47  \nl  
 ESOB316- 32     & 1 &10 &09 &06 &  -38 &24 &33 &SBab     &  1.925 &  13.53&       &         \nl  
 MRK26           & 1 &10 &11 &51 &   58 &53 &31 &Sc       &        &  15.99&       &         \nl  
 NGC3156         & 3 &10 &12 &41 &   03 &07 &50 &S0       & -2.451 &  13.08&   0.77&   0.26  \nl  
 NGC3256         & 3 &10 &27 &51 &  -43 &54 &20 &Sb       &  3.225 &  12.08&   0.64&  -0.07  \nl  
 MRK33           & 3 &10 &32 &31 &   54 &23 &56 &Irr      &  9.711 &  12.98&       &         \nl  
 NGC3311         & 1 &10 &36 &43 &  -27 &31 &41 &E-S0     & -3.429 &  12.42&   1.00&   0.57  \nl  
 NGC3310         & 4 &10 &38 &46 &   53 &30 &08 &SBbc     &  3.993 &  11.08&   0.35&  -0.43  \nl  
 NGC3351         & 2 &10 &43 &58 &   11 &42 &15 &SBb      &  3.077 &  10.39&   0.81&   0.19  \nl  
 NGC3353         & 2 &10 &45 &23 &   55 &57 &33 &Sb       &  3.004 &  13.22&   0.47&  -0.34  \nl  
 NGC3379         & 4 &10 &47 &50 &   12 &34 &57 &E        & -4.808 &  10.22&   0.96&   0.53  \nl  
 MRK153          & 1 &10 &49 &05 &   52 &19 &58 &Sc       &  4.900 &  14.98&   0.19&  -0.63  \nl  
 NGC3395         & 1 &10 &49 &49 &   32 &58 &51 &SBc      &  5.925 &  12.40&   0.34&  -0.23  \nl  
 NGC3396         & 1 &10 &49 &56 &   32 &59 &22 &SBm      &  9.435 &  12.48&       &         \nl  
 NGC3432         & 1 &10 &52 &31 &   36 &37 &08 &SBm      &  8.852 &  11.65&   0.42&  -0.36  \nl  
 MRK1267         & 2 &10 &53 &04 &   04 &37 &43 &E?       &        &  14.16&       &         \nl  
 ABCG1126        & 1 &10 &53 &50 &   16 &51 &00 &         &        &  15.12&       &         \nl  
 NGC3448         & 2 &10 &54 &39 &   54 &18 &24 &S0-a     &  0.200 &  12.41&   0.44&  -0.19  \nl  
 NGC3471         & 1 &10 &59 &09 &   61 &31 &51 &Sa       &  1.000 &  13.24&   0.71&   0.17  \nl  
 NGC3504         & 2 &11 &03 &11 &   27 &58 &25 &SBab     &  2.070 &  11.65&   0.73&   0.00  \nl  
 AOO APG 148     & 1 &11 &03 &54 &   40 &51 &00 &S?       &  9.200 &       &       &         \nl  
 MRK36           & 2 &11 &04 &58 &   29 &08 &22 &BCD      &        &  15.64&   0.28&  -0.66  \nl  
AOO CASG 816 W   & 1 &11 &12 &08 &   35 &52 &43 & compact &        &  17.50&       &         \nl  
 NGC3610         & 2 &11 &18 &26 &   58 &47 &14 &E        & -4.197 &  11.65&   0.87&   0.47  \nl   
 NGC3622         & 3 &11 &20 &13 &   67 &14 &27 &S?       &  3.800 &  13.65&   0.50&         \nl  
 NGC3640         & 2 &11 &21 &07 &   03 &14 &08 &E        & -4.846 &  11.30&   0.93&   0.53  \nl  
 NGC3642         & 2 &11 &22 &18 &   59 &04 &34 &Sbc      &  3.997 &  11.53&   0.50&         \nl  
 MRK170          & 2 &11 &26 &50 &   64 & 8 &16 &S?       &  2.700 &  15.03&       &         \nl  
 NGC3682         & 2 &11 &27 &43 &   66 &35 &25 &S0-a     & -0.037 &  13.26&   0.76&   0.04  \nl  
MCG +13-08-0058  & 1 &11 &28 &01 &   78 &59 &29 &pec      &        &  15.08&       &         \nl  
 NGC3690         & 8 &11 &28 &34 &   58 &33 &51 &SBm      &  8.754 &  11.71&       &         \nl  
 MRK178          & 2 &11 &33 &29 &   49 &14 &12 &Irr      &  9.933 &  14.45&   0.35&  -0.30  \nl
 NGC3738         & 3 &11 &35 &49 &   54 &31 &22 &Irr      &  9.783 &  11.98&   0.41&  -0.19  \nl  
 NGC3810         & 2 &11 &40 &58 &   11 &28 &17 &Sc       &  5.164 &  11.24&   0.58&  -0.06  \nl  
MCG +03-30-0066  & 1 &11 &43 &49 &   19 &58 &12 &Irr      &  9.702 &  14.10&   0.50&  -0.31  \nl  
 AOO ARP 248B    & 1 &11 &46 &45 &  -03 &50 &54 &SBbc     &  3.825 &  15.15&       &         \nl  
 NGC3894         & 1 &11 &48 &51 &   59 &25 &01 &E        & -4.090 &  12.63&   1.00&   0.59  \nl  
 NGC3991         & 4 &11 &57 &31 &   32 &20 &00 &SBd      &  7.900 &  13.52&   0.39&  -0.33  \nl  
 NGC3994         & 2 &11 &57 &36 &   32 &16 &44 &Sc       &  5.010 &  13.32&   0.62&  -0.14  \nl  
 NGC3995         & 2 &11 &57 &44 &   32 &17 &38 &SBm      &  8.833 &  12.69&   0.28&  -0.41  \nl  
 NGC4004         & 1 &11 &58 &05 &   27 &52 &38 &Irr      &  9.932 &  13.97&   0.44&  -0.27  \nl  
 AOO POX 36      & 1 &11 &58 &59 &  -19 &01 &36 &Scd      &  6.725 &  14.28&       &         \nl  
 AOO MKW 4       & 1 &12 &04 &27 &   01 &53 &48 &E        & -4.103 &  12.46&   1.00&         \nl  
AOO HE 1203-2644 & 2 &12 &05 &59 &  -27 &00 &54 &S0-a     & -1.000 &  14.98&       &         \nl  
 NGC4102         & 1 &12 &06 &23 &   52 &42 &41 &SBb      &  3.087 &  12.03&       &         \nl  
 NGC4111         & 4 &12 &07 &03 &   43 &04 &02 &S0-a     & -1.353 &  11.46&   0.89&   0.44  \nl  
 NGC4125         & 2 &12 &08 &07 &   65 &10 &22 &E        & -4.817 &  10.63&   0.94&   0.50  \nl  
 IC3017          & 1 &12 &09 &25 &   13 &34 &25 &S?       &  4.800 &  14.82&       &         \nl  
 AOO VCC 22      & 1 &12 &10 &24 &   13 &10 &13 &BCD?     &        &  16.13&       &         \nl  
 AOO VCC 24      & 2 &12 &10 &36 &   11 &45 &37 &BCD      &        &  15.10&       &         \nl  
 NGC4194         & 2 &12 &14 &10 &   54 &31 &39 &Irr      &  9.940 &  12.96&   0.55&  -0.20  \nl  
MCG +01-31-030   & 1 &12 &15 &19 &   05 &45 &42 &E        & -4.899 &  14.99&   0.39&  -0.47  \nl  
 NGC4214         & 8 &12 &15 &39 &   36 &19 &39 &Irr      &  9.810 &  10.13&   0.47&         \nl  
 AOO 1214-277    & 2 &12 &17 &21 &  -28 &02 &32 &EmLS     &        &       &       &         \nl  
 AOO 1214-28     & 1 &12 &17 &17 &  -28 &02 &33 &EmLS     &        &       &       &         \nl
 NGC4244         & 2 &12 &17 &30 &   37 &48 &27 &Sc       &  6.032 &  10.67&   0.50&         \nl  
 MRK49           & 1 &12 &19 &10 &   03 &51 &28 &E        & -4.924 &  14.29&   0.50&  -0.31  \nl  
 NGC4314         & 3 &12 &22 &32 &   29 &53 &47 &SBa      &  1.020 &  11.42&   0.86&   0.29  \nl  
 AOO VCC 562     & 1 &12 &22 &36 &   12 &09 &28 &BCD      &        &  16.43&       &  -0.61  \nl  
 NGC4321         & 7 &12 &22 &55 &   15 &49 &23 &SBbc     &  4.048 &  10.00&   0.70&  -0.01  \nl  
 NGC4350         & 5 &12 &23 &58 &   16 &41 &34 &S0       & -1.786 &  11.91&   0.95&   0.47  \nl  
 NGC4374         & 3 &12 &25 &04 &   12 &53 &15 &E        & -4.009 &  10.01&   0.99&   0.53  \nl  
 NGC4382         & 2 &12 &25 &25 &   18 &11 &27 &S0-a     & -1.332 &   9.94&   0.89&   0.42  \nl  
 NGC4385         & 4 &12 &25 &43 &   00 &34 &24 &S0-a     & -0.706 &  13.16&   0.69&   0.02  \nl  
 NGC4406         & 3 &12 &26 &12 &   12 &56 &49 &E        & -4.719 &   9.79&   0.94&   0.50  \nl  
 MRK209          & 3 &12 &26 &16 &   48 &29 &31 &Sm pec   &        &  14.65&       &         \nl  
 IC3370          & 1 &12 &27 &37 &  -39 &20 &18 &E        & -4.578 &  11.99&   0.98&   0.41  \nl  
 NGC4438         & 2 &12 &27 &45 &   13 &00 &36 &S0-a     &  0.500 &  10.89&   0.86&   0.35  \nl  
 NGC4449         &48 &12 &28 &11 &   44 &05 &40 &Irr      &  9.790 &   9.84&   0.41&  -0.34  \nl  
 NGC4472         &11 &12 &29 &46 &   07 &59 &58 &E        & -4.723 &   9.20&   0.96&   0.56  \nl  
 NGC4500         & 3 &12 &31 &22 &   57 &57 &52 &SBa      &  1.076 &  13.11&   0.61&  -0.07  \nl  
 NGC4494         & 1 &12 &31 &24 &   25 &46 &25 &E        & -4.817 &  10.68&   0.88&   0.45  \nl  
 NGC4552         & 6 &12 &35 &40 &   12 &33 &25 &E        & -4.611 &  10.68&   0.99&   0.56  \nl  
 NGC4566         & 1 &12 &36 &01 &   54 &13 &13 &Sbc      &  4.148 &  13.85&   0.75&  -0.01  \nl  
 NGC4564         & 1 &12 &36 &27 &   11 &26 &21 &E        & -4.717 &  11.86&   0.94&   0.47  \nl  
 NGC4621         & 3 &12 &42 &03 &   11 &38 &49 &E        & -4.766 &  10.67&   0.94&   0.48  \nl  
 NGC4649         & 4 &12 &43 &40 &   11 &32 &58 &E        & -4.580 &   9.71&   0.98&         \nl  
 MGC UGC 7905 S  & 1 &12 &43 &48 &   54 &53 &45 &Sab      &  1.913 &  14.06&       &         \nl  
 AOO HARO 33     & 3 &12 &44 &38 &   28 &28 &19 &Sc       &  5.516 &  14.74&       &         \nl  
 NGC4650A        & 1 &12 &44 &50 &  -40 &42 &54 &S0-a     &  0.000 &  13.92&       &         \nl  
 NGC4670         & 3 &12 &45 &17 &   27 &07 &34 &S0-a     &  0.301 &  13.07&   0.41&  -0.47  \nl  
MCG +02-33-0012  & 1 &12 &46 &05 &   08 &28 &31 &BCD      &        &  14.78&       &         \nl  
AOO ANON 1244-53 & 2 &12 &47 &38 &  -53 &33 &08 &E?       & -0.500 &  14.80&       &         \nl  
 NGC4697         & 4 &12 &48 &36 &  -05 &48 &02 &E        & -4.789 &  10.13&   0.92&   0.39  \nl  
 NGC4696         & 2 &12 &48 &48 &  -41 &18 &39 &E        &  3.868 &  11.54&       &         \nl  
AOO TOL 1247-232 & 1 &12 &50 &19 &  -23 &33 &57 &EmLS     &        &       &       &         \nl  
 NGC4736         &11 &12 &50 &54 &   41 &07 &10 &Sb       &  2.519 &   8.74&   0.75&   0.16  \nl  
 NGC4742         & 2 &12 &51 &48 &  -10 &27 &18 &E        & -4.788 &  12.08&   0.81&   0.31  \nl  
 NGC4762         & 4 &12 &52 &56 &   11 &13 &48 &S0       & -1.826 &  11.11&   0.87&   0.41  \nl  
 NGC4774         & 1 &12 &53 &07 &   36 &49 &07 & Sd      &  7.867 &  14.81&   0.48&  -0.09  \nl  
 MRK54           & 5 &12 &56 &56 &   32 &26 &55 &Sc       &  5.056 &  15.29&   0.28&         \nl  
 NGC4853         & 4 &12 &58 &35 &   27 &35 &50 &E-S0     & -2.951 &  14.40&   0.81&   0.24  \nl  
 NGC4861         & 6 &12 &59 &03 &   34 &51 &38 &SBm      &  8.932 &  14.10&       &         \nl  
 NGC4874         & 1 &12 &59 &36 &   27 &57 &44 &E        & -3.634 &  12.71&   0.95&   0.50  \nl  
 NGC4889         & 2 &13 &00 &08 &   27 &58 &45 &E        & -4.301 &  12.53&   1.04&   0.52  \nl  
 QSO 1300+361    & 1 &13 &03 &03 &   35 &51 &29 &         &        &  18.00&       &         \nl
 AOO POX 120     & 1 &13 &06 &42 &  -12 &04 &22 &EmLS     &        &  15.70&       &         \nl  
 AOO POX 124     & 1 &13 &07 &26 &  -13 &11 &01 &EmLS     &        &  15.51&       &         \nl  
 NGC5018         & 2 &13 &13 &01 &  -19 &31 &12 &E        & -4.548 &  11.66&   0.93&   0.48  \nl  
 MCG +07-27-0052 & 1 &13 &14 &10 &   39 &08 &51 &         &        &  15.58&       &         \nl  
 MRK450          & 2 &13 &14 &48 &   34 &52 &44 &Irr      &  9.846 &  14.33&   0.56&  -0.43  \nl  
 NGC5044         & 1 &13 &15 &24 &  -16 &23 &09 &E        & -4.778 &  11.69&   1.00&   0.55  \nl  
 MGC UGC 8335 N  & 1 &13 &15 &35 &   62 &07 &27 &Sab      &  1.700 &  14.62&       &         \nl  
 NGC5055         & 1 &13 &15 &49 &   42 &02 &06 &Sbc      &  3.971 &   9.18&   0.73&         \nl  
 NGC5102         & 8 &13 &21 &58 &  -36 &37 &47 &E-S0     & -2.972 &  10.28&   0.73&   0.23  \nl  
 NGC5122         & 2 &13 &24 &15 &  -10 &39 &16 &S?       &  2.700 &  14.10&       &         \nl  
 AOO POX 186     & 3 &13 &25 &51 &  -11 &37 &35 &EmLS     &        &  17.00&       &         \nl  
 MRK66           & 1 &13 &25 &54 &   57 &15 &05 &BCG      &        &  15.05&       &         \nl  
 NGC5204         & 1 &13 &29 &36 &   58 &25 &04 &Sm       &  8.851 &  11.66&   0.41&  -0.33  \nl  
 NGC5195         & 1 &13 &29 &59 &   47 &16 &05 &S0-a     &  0.064 &  10.26&   0.90&   0.31  \nl  
NGC5236          &23 &13 &37 &00 &  -29 &52 &04 &SBc      &  5.028 &   7.92&   0.67&   0.04  \nl  
ESOB383-G44      & 2 &13 &37 &27 &  -33 &00 &22 &Scd      &  6.815 &  14.05&       &         \nl  
 NGC5253         &19 &13 &39 &56 &  -31 &38 &41 &S?       &  7.689 &  10.77&   0.44&  -0.23  \nl  
 MRK67           & 1 &13 &41 &56 &   30 &31 &11 &BCD      &        &  16.36&       &         \nl  
 NGC5266         & 1 &13 &43 &02 &  -48 &10 &12 &E-S0     & -2.589 &  12.13&       &         \nl  
 NGC5291         & 1 &13 &47 &24 &  -30 &24 &27 &E        & -3.940 &  13.64&       &         \nl  
 ABCG1795        & 7 &13 &48 &52 &   26 &35 &35 & cD-S0   &        &  15.20&   1.00&    .44  \nl  
AOO Z 13502+0022 & 1 &13 &52 &44 &   00 &07 &51 &Sm       &  9.000 &  15.43&   0.65&         \nl  
 MCG +04-33-038  & 1 &14 &01 &09 &   21 &14 &15 &S?       &  3.400 &  15.04&       &         \nl  
 NGC5398         & 3 &14 &01 &21 &  -33 &03 &46 &SBd      &  7.836 &  13.84&       &         \nl  
 AOO TOL 41      & 1 &14 &02 &59 &  -30 &14 &25 &EmLS     &        &  18.00&       &         \nl  
 NGC5457         & 6 &14 &03 &13 &   54 &21 &03 &SBc      &  5.944 &   8.20&   0.45&         \nl  
 NGC5408         & 5 &14 &03 &21 &  -41 &22 &35 &Irr      &  9.918 &  12.21&   0.56&  -0.33  \nl  
 AOO  1401+114   & 1 &14 &03 &24 &   11 &09 &14 &         &        &       &       &         \nl
MCG +02-36-0024  & 1 &14 &03 &27 &   09 &27 &57 &Sbc      &  4.005 &  14.67&       &         \nl  
 NGC5461         & 7 &14 &03 &42 &   54 &18 &59 &         &        &       &       &         \nl  
 NGC5471         &13 &14 &04 &28 &   54 &23 &49 &         &        &       &       &         \nl  
AOO PKS 1404-267 & 1 &14 &07 &31 &  -27 &01 &02 &E-S0     &  0.008 &  13.73&       &         \nl 
 NGC5670         & 2 &14 &35 &37 &  -45 &57 &58 &S0       & -2.010 &  12.96&   1.00&   0.50  \nl  
 IC4448          & 1 &14 &40 &28 &  -78 &48 &37 & SBd     &  7.670 &  14.06&   0.58&  -0.14  \nl  
 MRK288          & 1 &14 &50 &47 &   73 &49 &24 &S?       &  4.900 &  15.70&       &         \nl  
MCG +06-33-0004  & 3 &14 &50 &57 &   35 &34 &17 &Sd       &  8.000 &  14.77&   0.26&  -0.72  \nl  
 ABCG1991        & 2 &14 &54 &32 &   18 &38 &24 &E        & -4.817 &  15.29&       &         \nl  
 NGC5846         & 1 &15 &06 &29 &   01 &36 &25 &E        & -4.677 &  10.98&   1.01&   0.48  \nl  
 AOO CASEG 657   & 1 &15 &12 &13 &   47 &16 &31 &         &        &  16.30&       &         \nl  
 QSO 1514+072    & 2 &15 &16 &44 &   07 &01 &16 &E?       & -2.025 &  14.52&       &         \nl  
 MRK487          & 1 &15 &37 &04 &   55 &15 &47 &E        & -5.000 &  15.46&   0.50&  -0.41  \nl  
 NGC5996         & 2 &15 &46 &59 &   17 &53 &08 &Sc       &  5.339 &  13.01&   0.42&  -0.25  \nl  
 MRK492          & 1 &15 &58 &44 &   26 &48 &51 &E-S0     & -2.633 &  14.86&   0.83&   0.22  \nl   
 NGC6052         & 3 &16 &05 &13 &   20 &32 &38 &Sc       &  5.058 &  13.45&   0.44&  -0.43  \nl  
 NGC6062         & 4 &16 &06 &23 &   19 &46 &44 &SBbc     &  3.833 &  14.23&   0.58&  -0.17  \nl  
 NGC6090         & 1 &16 &11 &40 &   52 &27 &21 &Sab      &  1.700 &  14.49&       &         \nl  
 NGC6166         & 5 &16 &28 &38 &   39 &33 &04 &E        & -4.327 &  12.86&   1.00&         \nl  
 MRK499          & 1 &16 &48 &24 &   48 &42 &23 &Im       &        &  14.60&       &         \nl  
 NGC6340         & 1 &17 &10 &26 &   72 &18 &22 &S0-a     &  0.396 &  11.95&   0.87&         \nl  
AOO AM 1724-622  & 1 &17 &29 &09 &  -62 &26 &43 &S0-a     & -1.000 &  12.68&       &         \nl  
 NGC6500         & 1 &17 &55 &59 &   18 &20 &26 &Sab      &  1.621 &  13.01&       &         \nl  
 AOO FRL 44      & 1 &18 &13 &39 &  -57 &43 &58 &S? pec   &        &       &       &         \nl 
ESOB338-IG4      &15 &19 &27 &58 &  -41 &34 &28 &pec      &        &  13.42&   0.29&  -0.41  \nl  
NGC6822          & 7 &19 &44 &58 &  -14 &48 &11 &Irr      &  9.777 &   9.33&       &         \nl  
ESOB185-IG13     & 1 &19 &45 &01 &  -54 &15 &03 &compact  &        &  15.00&   0.19&  -0.17  \nl  
 NGC6868         & 1 &20 &09 &54 &  -48 &22 &44 &E        & -4.447 &  11.61&   1.01&   0.62  \nl  
AOO AM 2020-504  & 4 &20 &23 &56 &  -50 &39 &06 &pec ring ? &        &       &       &         \nl  
 ESOB462-IG20    & 1 &20 &26 &57 &  -29 &07 &06 &E        & -4.000 &  14.60&       &         \nl  
 NGC6946         & 2 &20 &34 &52 &   60 &09 &15 &SBc      &  5.938 &   9.58&   0.81&         \nl  
 ESOB400-G43     & 1 &20 &37 &42 &  -35 &29 &11 &         &        &  14.28&       &         \nl  
 NGC7083         & 2 &21 &35 &45 &  -63 &54 &17 &Sbc      &  4.006 &  11.89&   0.65&   0.02  \nl  
 NGC7173         & 1 &22 &02 &04 &  -31 &58 &25 &E        & -4.173 &  13.14&   0.93&   0.47  \nl  
 NGC7176         & 1 &22 &02 &09 &  -31 &59 &25 &E        & -4.644 &  12.42&       &         \nl  
 NGC7196         & 2 &22 &05 &55 &  -50 &07 &11 &E        & -4.764 &  12.63&   0.94&   0.47  \nl  
 NGC7233         & 3 &22 &15 &49 &  -45 &50 &51 &S0-a     &  0.371 &  13.53&   0.62&  -0.01  \nl  
 NGC7250         & 3 &22 &18 &18 &   40 &33 &45 &Sd       &  7.960 &  13.20&   0.64&  -0.04  \nl  
 NGC7252         & 2 &22 &20 &45 &  -24 &40 &41 &S0       & -2.026 &  13.47&   0.67&   0.20  \nl  
 NGC7412         & 5 &22 &55 &45 &  -42 &38 &28 &SBb      &  3.327 &  11.91&   0.53&   0.02  \nl  
 NGC7513         & 3 &23 &13 &13 &  -28 &21 &34 &SBb      &  3.118 &  12.71&   0.76&   0.12  \nl  
MCG -07-47-023   & 2 &23 &13 &59 &  -42 &43 &39 &E        & -3.873 &  15.51&       &         \nl  
 NGC7552         & 4 &23 &16 &11 &  -42 &35 &01 &SBab     &  2.385 &  11.20&   0.69&   0.10  \nl  
 NGC7609 B       & 2 &23 &19 &31 &   09 &30 &10 &Sd       &  8.462 &  16.00&       &         \nl  
 NGC7648         & 1 &23 &23 &54 &   09 &40 &04 &S0       & -1.941 &  13.77&   0.81&   0.28  \nl  
 ABCG2597        & 1 &23 &25 &20 &  -12 &07 &27 &E        &        &  16.32&       &         \nl  
 NGC7673         & 2 &23 &27 &42 &   23 &35 &24 &Sc       &  4.950 &  13.10&   0.41&  -0.33  \nl  
AOO Z 2327.6+251 & 2 &23 &30 &09 &   25 &31 &43 &SB+S0    &        &  15.06&       &         \nl  
 NGC7714         & 5 &23 &36 &15 &   02 &09 &18 &SBb      &  3.106 &  12.98&   0.52&  -0.44  \nl  
 ABCG2626        & 1 &23 &36 &30 &   21 &08 &48 &Sc       &  5.056 &  15.31&       &         \nl  
 ABCG2634        & 1 &23 &38 &30 &   27 &01 &51 &E        & -4.100 &  13.47&   1.05&   0.49  \nl  
 NGC7771         & 2 &23 &51 &25 &   20 &06 &49 &SBa      &  1.050 &  13.07&   0.83&   0.34  \nl  
 NGC7793         & 6 &23 &57 &49 &  -32 &35 &24 &Scd      &  7.429 &   9.70&       &         \nl  

\enddata

\tablecomments{The positional information is for J2000. It is presented in columns 3, 4, 5 
(Right Ascension), and 6, 7, 8 (Declination).}

\end{deluxetable}
\begin{deluxetable}{cccccrcc}
\tablecaption{Revised aperture coordinates}
\small
\tablehead{\colhead{IUE ID} & \colhead{ Image ID} & \colhead{h } & \colhead{m} & \colhead{s} & 
\colhead{$^{\circ}$} & \colhead{'} & \colhead{"}  }
\startdata

 IC1586            & SWP27190 & 00 & 45 & 17.3 & 22 & 06 & 05.2  \nl
 NGC584            & SWP24973 & 01 & 28 & 50.8 &-07 & 07 & 28.6  \nl
 MRK600            & SWP22347 & 02 & 48 & 27.2 & 04 & 14 & 53.9  \nl
 NGC1404           & SWP09189 & 03 & 36 & 57.5 &-35 & 45 & 19.9  \nl
 NGC1404           & LWR07958 & 03 & 36 & 57.6 &-35 & 45 & 20.1  \nl
 MRK7              & SWP11139 & 07 & 22 & 20.0 & 72 & 40 & 33.5  \nl
 NGC3125           & LWR09894 & 10 & 04 & 18.3 &-29 & 41 & 26.5  \nl
 MRK153            & SWP17314 & 10 & 46 & 04.5 & 52 & 36 & 02.3  \nl
 MRK1267           & LWP01769 & 10 & 50 & 28.3 & 04 & 53 & 51.8  \nl 
 MCG +03-30-0066   & SWP33212 & 11 & 41 & 13.4 & 20 & 14 & 36.7  \nl
 NGC3995           & LWR12534 & 11 & 55 & 09.9 & 32 & 34 & 23.6  \nl
 NGC4194           & SWP22784 & 12 & 11 & 41.6 & 54 & 48 & 16.9  \nl
 NGC4194           & LWP03412 & 12 & 11 & 41.7 & 54 & 48 & 17.7  \nl
 NGC4244           & SWP20476 & 12 & 15 & 04.3 & 38 & 04 & 38.9  \nl
 MRK209            & LWP12696 & 12 & 23 & 50.7 & 48 & 46 & 05.7  \nl
 NGC4566           & LWR14420 & 12 & 33 & 40.3 & 54 & 29 & 50.7  \nl
 NGC5044           & SWP41960 & 13 & 12 & 43.6 &-16 & 07 & 42.9  \nl
 MRK66             & SWP24384 & 13 & 23 & 58.5 & 57 & 30 & 54.2  \nl
 MCG +02-36-0024   & SWP32987 & 14 & 00 & 59.2 & 09 & 42 & 39.3  \nl
 MCG +02-36-0004   & LWR09826 & 14 & 48 & 54.9 & 35 & 46 & 34.1  \nl
 NGC7250           & SWP11227 & 22 & 16 & 09.1 & 40 & 18 & 43.4  \nl
 NGC7250           & LWR09846 & 22 & 16 & 09.0 & 40 & 18 & 42.6  \nl
 NGC7673           & LWR09760 & 23 & 25 & 11.8 & 23 & 18 & 59.4  \nl
 
\enddata

\tablecomments{The positional information is for J1950, in order to match
the coordinate system of the INES header. For the following
images no information on the guide star was stored: SWP04207 (NGC520), LWR14561
 (NGC2784), LWR13025 (NGC4125), LWR10631 (NGC4500), SWP07918 (NGC5471), 
SWP07936 (MCG +06-33-0004), SWP07303 (NGC7673).}

\end{deluxetable}

\end{document}